\documentclass{article}

\PassOptionsToPackage{numbers,sort&compress}{natbib}
\usepackage[preprint]{neurips_2026}

\usepackage[utf8]{inputenc} 
\usepackage[T1]{fontenc}    
\usepackage{hyperref}       
\usepackage{url}            
\usepackage{booktabs}       
\usepackage{amsfonts}       
\usepackage{nicefrac}       
\usepackage{microtype}      
\usepackage{xcolor}         
\usepackage{amsmath}
\usepackage{amssymb}
\usepackage{graphicx}
\usepackage{tabularx}
\usepackage{makecell}
\usepackage{listings}
\usepackage{placeins}

\usepackage[acronym, automake]{glossaries}
\usepackage{glossary-mcols}

\lstdefinestyle{promptstyle}{%
  basicstyle=\ttfamily\footnotesize,%
  breaklines=true,%
  breakatwhitespace=false,%
  columns=fullflexible,%
  keepspaces=true,%
  showstringspaces=false,%
  upquote=true,%
  frame=single,%
  framerule=0.4pt,%
  rulecolor=\color{black!30},%
  backgroundcolor=\color{black!3},%
  xleftmargin=0.5em,%
  xrightmargin=0.5em,%
  aboveskip=0.6em,%
  belowskip=0.6em,%
  literate={--}{{-{}-}}2 {<=}{{<=}}2 {>=}{{>=}}2,%
}
\setacronymstyle{long-short}
\makeglossaries

\setglossarystyle{mcolindex}

\title{ARA: Agentic Reproducibility Assessment\\For Scalable Support Of Scientific Peer-Review}

\author{%
  Kevin Riehl$^1$ \\
  ETH Zürich, IVT \& \\
  Agentic Systems Lab (ASL) \\
  \texttt{kriehl@ethz.ch} \\
  \And
  Andres L. Marin$^1$ \\
  European Commission, Joint Research Centre \& \\
  University of Konstanz \\
  \texttt{E-Mail: andres.laverde-marin@ec.europa.eu} \\
  \And
  Nikiforos Zacharof \\
  Ideas Forward \\
  \texttt{nikiforos.zacharof@ideasforward.com} \\
  \And
  Fan Wu \\
  ETH Zürich, ASL \\
  \texttt{E-Mail: fanwu@ethz.ch} \\
  \And
  Patrick Langer \\
  ETH Zürich, ASL \\
  \texttt{E-Mail: planger@ethz.ch} \\
  \And
  Robert Jakob \\
  ETH Zürich, ASL \\
  \texttt{E-Mail: rjakob@ethz.ch} \\
  \\  \And
  Anastasios Kouvelas \\
  ETH Zürich, IVT \\
  \texttt{kouvelas@ethz.ch} \\
  \And
  Georgios Fontaras \\
  European Commission, Joint Research Centre \\
  \texttt{georgios.fontaras@ec.europa.eu} \\
  \And
  Michail A. Makridis \\
  ETH Zürich, IVT \\
  \texttt{mmakridis@ethz.ch} \\
}

\begin{document}

\maketitle

\begin{abstract}
Scientific peer review increasingly struggles to assess reproducibility at the scale and complexity of modern research output.
Evaluating reproducibility requires reconstructing experimental dependencies, methodological choices, data flows, and result-generating procedures, which often exceeds what human reviewers can provide given the growing number of submissions and time constraints.
Agentic Reproducibility Assessment (ARA) formalizes reproducibility assessment as a structured reasoning task over scientific documents.
Given a paper, ARA extracts a directed workflow graph linking sources, methods, experiments, and outputs, then evaluates its reconstructability using structural and content-based scores for reproducibility assessments.
Experiments on 213 ReScience C articles -- the largest cross-domain benchmark of human-validated computational reproducibility studies considered to date -- demonstrate ARA's generalizability and consistent workflow reconstruction and assessment across LLMs, model temperatures, and scientific domains.
ARA consistently achieves 61\% accuracy on three reproducibility assessment benchmarks, 
highlighting its potential to complement human review at scale and enabling next-generation peer review.
\end{abstract}

\section{Introduction}

Scientific publication output has grown exponentially across disciplines over the past decades, driven by expanding global research communities, increasing specialization of venues, and strong incentive structures tied to publication productivity~\citep{parchomovsky2000publish,szalay2006science}. 
More recently, the emergence of large language models (LLMs) as writing assistants has further accelerated manuscript production, raising concerns about declining signal-to-noise ratios in the scientific literature~\citep{mosca2023distinguishing,cheng2024have}. 
While increased output reflects healthy research activity, it also places substantial pressure on traditional quality assurance mechanisms such as peer review, the backbone of modern science and  an essential part of the knowledge production process, which was designed for slower and smaller publishing ecosystems~\citep{zhou2024llm}.
Today, reviewers face growing workloads, limited time, and heterogeneous evaluation standards, making it increasingly difficult to maintain consistent quality, rigorous assessment, and verifiable reproducibility across submissions~\citep{ellison2011peer}. 
As a result, scalable support mechanisms for evaluating the quality and reliability of scientific claims are becoming an urgent requirement~\citep{wei2025ai}.

Among the various dimensions of scientific quality, reproducibility occupies a central role as a necessary condition for validating scientific contributions~\citep{popperDE,popperEN}. 
However, large-scale meta-analyses across multiple disciplines including psychology~\citep{anvari2018replicability}, medicine and biomedical sciences~\citep{errington2014open,stupple2019reproducibility,miyakawa2020no}, social and behavioral sciences~\citep{fivsar2024reproducibility,tyner2026investigating}, and computer science~\citep{hutson2018artificial,riehl2025revisiting} have documented a substantial reproducibility crisis~\citep{baker2016reproducibility,fanelli2018science}.
Many published studies omit essential implementation details such as parameter settings, preprocessing steps, or access to datasets and source code, limiting the ability of researchers to independently verify reported findings and to assess the generalizability of the resulting knowledge~\citep{stark2018before}.
Importantly, assessing reproducibility is not equivalent to evaluating narrative clarity or novelty; it requires reconstructing procedural dependencies and experimental pipelines, often across multiple methodological layers. 
Such efforts exceed what volunteer peer reviewers can realistically provide during standard evaluation cycles. 
Consequently, reproducibility assessments remain underrepresented in standard peer-review evaluation workflows, and difficult to scale relative to the volume of published research~\citep{national2019reproducibility}.

Recent advances in LLMs and agentic reasoning systems suggest a new opportunity to address this gap. 
Modern agent pipelines are capable of operating over long technical documents, extracting structured procedural information, identifying dependencies between experimental components, and performing multi-step reasoning tasks across heterogeneous textual evidence~\citep{acharya2025agentic}. 
Unlike manual review processes, these systems scale naturally with corpus size and can be deployed consistently across large collections of scientific articles. 
This creates the possibility of operationalizing reproducibility assessment itself as a structured reasoning problem that can be partially automated~\citep{li2025generation}.
Rather than replacing human replication efforts, agentic systems may provide scalable diagnostic support for identifying missing reporting elements, incomplete workflows, and reproducibility bottlenecks in scientific publications.
This suggests that reproducibility assessment may be tractable as a document-level reasoning task rather than requiring full experimental replication~\citep{jin2024agentreview,thakkar2025can}.
Previous work has demonstrated the potential of agentic systems for peer-review~\citep{liang2024can,hossain2025llms,zhuang2025large}, to assess reproducibility ~\citep{hu2025repro,nguyen2026replicatorbench}, or even to reproduce papers~\citep{lupidi2026airsbenchsuitetasksfrontier,kim2025reproduction,seo2025paper2code}. However, these preliminary approaches are limited to specific domains, rely on simple techniques such as document-parsing and checklist-based approaches rather than reasoning over the structure of the paper, and therefore lack generalizability. 

\begin{figure}[!hb]
    \centering
    \includegraphics[width=0.8\linewidth]{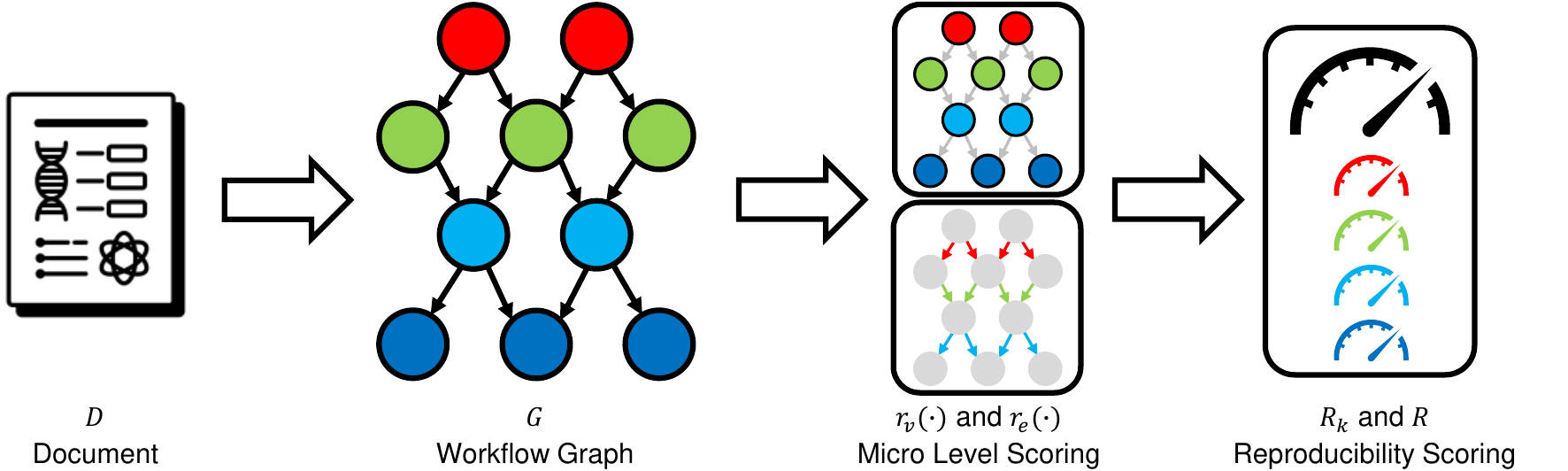}
    \caption{
\textbf{Agentic Reproducibility Assessment Pipeline (\textit{ARA}).}
First, a given scientific paper (resp. document) $D$ is transformed into a directed workflow graph $G$, comprising four types of nodes (sources, methods, experiments, sinks). 
Second, the workflow graph's reconstructability is projected on micro-level assessments of reproducibility (node-by-node) $r(\cdot)$.
Third, the micro-level assessments are aggregated to reproducibility scores $R_k$ and $R$.
    }\label{fig:ara_pipeline}
\end{figure}

In this work, we investigate whether and how reproducibility assessment can be formalized as a structured reasoning task executable by agentic AI systems. 
We introduce \texttt{Agentic Reproducibility Assessment (ARA)}, an LLM-agent pipeline that converts scientific papers into workflow graphs of their required reproduction components and projects them onto multi-dimensional reproducibility scores.
We ask: \textbf{(RQ1)} can one methodology consistently assess reproducibility across scientific domains? \textbf{(RQ2)} does it generalize across large paper datasets, LLMs, and temperatures? \textbf{(RQ3)} can it be trusted as a high-level tool that strengthens peer review?
To answer these research questions, we evaluate the proposed framework using 213 scientific publications (largest dataset to date) from \textit{ReScience C}, a journal dedicated to computational reproducibility studies across a multitude of scientific disciplines~\citep{rougier2018rescience}. 
Papers in this dataset provide structured documentation of human replication attempts and therefore constitute a uniquely suitable benchmark dataset for assessing automated reproducibility diagnostics versus human assessments.
Using this dataset as a benchmark, we compare agent-generated reproducibility assessments with verified reproduction workflows and reporting structures.
The experiments are complemented by consistency checks across different LLMs, temperatures, and further benchmark datasets.
Our results demonstrate the domain-agnostic capabilities of \texttt{ARA}, and that the proposed agentic reproducibility analysis captures meaningful signals aligned with real-world reproducibility requirements beyond mere checklist-based document parsing. 
\section{Related Work}

Prior work on reproducibility assessment spans checklist-based evaluation frameworks, automated reconstruction of experimental workflows from scientific papers, and recent efforts to apply LLM agents to peer review and replication support. 
These developments suggest that scientific reliability assessment can increasingly be approached as a structured document-level reasoning task. 
However, existing approaches remain largely domain-specific, depend on predefined reporting metrics, or assume access to replication packages beyond the publication itself. 
In the following, we review related work across these directions and position \texttt{ARA} as a domain-agnostic workflow reconstruction framework for scalable reproducibility assessment directly from scientific papers.


\paragraph{Reproducibility Assessment.}
Reproducibility is a foundational principle of modern scientific practice because it enables independent verification of empirical claims~\citep{riehl2025revisiting} and supports the falsification criterion central to Popperian critical rationalism~\citep{popperDE,popperEN}. 
Prior work conceptualizes reproducibility as a multi-dimensional construct encompassing the availability of supplementary materials and digital research artifacts~\citep{riehl2025revisiting,stagge2019assessing}, the reliability and repeatability of methodological procedures~\citep{downing2004reliability,bazzocchi2011accuracy,crowley2016critical,bunting2019practical}, transparency in statistical reporting~\citep{hung2020statistical,held2024assessment}, and the use of standardized benchmarks and harmonized experimental protocols~\citep{crowley2016critical,arroyo2022systematic}. 
Additional work emphasizes systematic reporting of data quality and experimental assumptions, as well as the development of quantitative reproducibility indicators beyond qualitative reporting assessments~\citep{crowley2016critical}. 
Empirical meta-research further documents persistent barriers to reproducibility assessment, including missing descriptive statistics, unavailable datasets, insufficient methodological detail, and limited author responsiveness to clarification requests~\cite{errington2014open}. 
Despite these advances, most reproducibility assessment frameworks remain domain-specific and rely on checklist-based criteria that capture only limited aspects of methodological completeness, limiting their generalizability across disciplines.
\paragraph{Agentic \& LLM-assisted Paper Reproduction.}
Recent advances in LLMs and agentic systems have enabled automated pipelines that translate scientific papers into executable computational workflows, particularly in computer science~\citep{seo2025paper2code,li2025deepcode}, astrophysics~\citep{ye2025replicationbench}, machine learning~\citep{starace2025paperbench}, and quantitative science~\citep{kubota2026llm}, where experiments rely on standardized datasets, statistical models, and software-based implementations~\citep{kubota2026llm}. 
A growing line of work studies document-to-codebase synthesis~\citep{li2025deepcode}, reconstructing repository-level implementations from textual method descriptions through hierarchical information-flow optimization~\citep{li2025deepcode}. 
Related systems such as \textit{PaperCoder} (\textit{Paper2Code})~\citep{seo2025paper2code} further demonstrate that agent pipelines can plan architectures, identify file dependencies, and generate modular implementations from scientific documents. 
Benchmarks including \textit{AutoExperiment}~\citep{kim2025reproduction}, \textit{ReplicationBench}~\citep{ye2025replicationbench}, and \textit{PaperBench}~\citep{starace2025paperbench} evaluate agents’ ability to synthesize experimental code and reproduce reported results. 
Recent work further shows that LLM-based systems can replicate statistical analyses from quantitative social science papers and detect inconsistencies in reported results, supporting their integration into extended scientific workflows~\citep{kubota2026llm,kubota2026llm2,ye2025replicationbench}. 
In contrast to these preliminary, domain-specific approaches, which aim to reconstruct or execute experiments directly, \texttt{ARA} models reproducibility assessment as a structured reasoning task that evaluates methodological dependency completeness without requiring full experimental replication.

\paragraph{Agentic \& LLM-assisted Scholarly Peer Review.}
Recent work explores the use of large language models as assistants in scholarly peer review to address growing submission volumes and reviewer workload. 
Empirical studies show that LLM-generated comments substantially overlap with human feedback and are often perceived as similarly helpful by authors, supporting their feasibility for early-stage manuscript evaluation~\citep{liang2024can}. 
Automated reviewing systems have further been proposed for rapid quality screening, meta-review synthesis, and editorial decision support at scale~\citep{tyser2024ai,hossain2025llms}. 
Alternative paradigms evaluate manuscripts through pairwise comparison with existing literature rather than absolute scoring, improving impact identification but introducing risks of topical and institutional bias~\citep{zhang2025replication}. 
At the same time, technical evaluations highlight reliability limitations on longer manuscripts and the need for human oversight~\citep{zhou2024llm}. 
Survey evidence indicates that LLM-assisted reviewing is already widespread, with approximately 19\% of reviewers reporting usage in 2025 and more than 50\% in 2026, raising concerns about transparency and the social contract of peer review~\citep{naddaf2025ai,naddaf2026more}. 
More broadly, increasing reliance on LLM-mediated reviewing may introduce feedback loops in which scientific writing and evaluation co-evolve with model-generated outputs, raising governance challenges for AI-assisted scholarly communication~\citep{zhuang2025large}.

\paragraph{Agentic \& LLM-assisted Reproducibility Assessment.}

Recent work explores the use of LLMs for automated reproducibility assessment, typically through checklist-based scoring frameworks or domain-specific evaluation pipelines. 
For example, \textit{Reproscreener} evaluates computational reproducibility signals in machine learning preprints using structured reporting metrics derived from publication standards~\citep{bhaskar2024reproscreener}, while \textit{Paper-snitch}~\citep{santoli2024} and \textit{RECAP}~\citep{da2026assessing} automate artifact availability and checklist-style transparency screening in specialized conference corpora. 
Similarly, \textit{Auto-METRICS}~\citep{de2025auto} demonstrates that LLMs can assist methodological quality assessment in radiomics research using standardized domain-specific scoring protocols. 
More recent agent-based benchmarks such as \textit{ReplicatorBench}~\citep{nguyen2026replicatorbench} evaluate end-to-end replication workflows across stages of experimental reconstruction and interpretation, but remain restricted to social and behavioral science settings. 
The closest related work, \textit{REPRO-BENCH} and \textit{REPRO-AGENT}~\citep{hu2025repro}, provides early evidence of agentic capabilities for reproducibility assessment in social science (given access to code, data, and available reproduction packages).
In contrast, the proposed framework \texttt{ARA} models reproducibility assessment as a document-level workflow reconstruction problem, enabling domain-agnostic reproducibility assessments across substantially larger and more heterogeneous corpora of scientific studies.


\section{Methods}

The proposed domain-agnostic pipeline \texttt{ARA} models reproducibility assessment as structured reasoning over evidence-generation workflows extracted from publication text. 
The framework is presented in Figure~\ref{fig:ara_pipeline}, and operates in three steps which are described in the following subsections, followed by the presentation of the \textit{ReScience~C} dataset used for evaluation.

\subsection{Workflow Graph Construction}

Given a scientific paper represented as a document $D$, \texttt{ARA} constructs a directed workflow graph $G = (V, E)$.
Each node $v\in V$ corresponds to a methodological component (e.g., dataset, preprocessing step, model, parameterization, evaluation procedure, or result), and each edge $e\in E$ represents a dependency relation between components.
\texttt{ARA} interprets reproducibility as the reconstructability of workflow graph dependency paths and links (across four stages sources, methods, experiments, and sinks) rather than the mere availability of external artifacts.
Reproducibility is evaluated as the degree to which this pathway can be reconstructed from the information contained in the paper alone, without requiring access to external artifacts such as source code or replication packages.
Nodes $V$ are partitioned into four, domain-agnostic workflow stages $\mathcal{K}$: 
\begin{equation}
    V = V_{\text{sources}} \cup V_{\text{methods}} \cup V_{\text{experiments}} \cup V_{\text{sinks}}
\end{equation}
where each stage captures distinct components of the evidence-generation pipeline:
\begin{itemize}
    \item \textbf{Sources.}
    This dimension evaluates whether datasets, assumptions, and research questions are sufficiently specified to initialize the computational workflow.
    \item \textbf{Methods.}
    This dimension evaluates whether intermediate transformations, algorithms, variables, parameterizations, software \& hardware resources, and modeling steps are described clearly enough to permit independent implementation.
    \item \textbf{Experiments.}
    This dimension evaluates whether evaluation protocols, baselines, runtime conditions, software \& hardware resources, and experimental configurations are reproducible.
    \item \textbf{Sinks.}
    This dimension evaluates whether result-relevant figures, tables, and reported conclusions can be traced to explicitly described workflow steps and supporting evidence.
\end{itemize}

Workflow graphs are constructed using an LLM-based agentic extraction pipeline using the full paper as a context. 
The agent issues stage-specific structured queries to extract datasets, methodological steps, experimental configurations, parameters, and reported results. 
Responses are schema-constrained and grounded in literal source quotes to ensure traceability. 
Extracted components are mapped to workflow nodes and assigned to one of four stages (sources, methods, experiments, sinks), and stage-specific attributes required for reconstructability scoring are derived automatically. 
An exemplary workflow graph construction can be found in Appendix~\ref{appendix:ex_workflow}.
Detailed prompt templates, system instructions, and attribute definitions are provided in Appendix~\ref{appendix:prompts}.



\subsection{Micro-Level Scoring}

Reconstructability of individual nodes is assessed using an ordinal scoring function $r(\cdot)$. 
We adopt a four-level scale because it provides an interpretable representation of reconstruction feasibility while remaining comparable to prior reproducibility assessment protocols~\citep{stagge2019assessing,arroyo2022systematic,brodeur2024mass,held2024assessment,hu2025repro}. 
Unlike binary labels, the ordinal scale captures intermediate levels of specification without requiring execution-based verification, while avoiding excessive granularity that may reduce annotation reliability. 
Specifically, $1$ denotes missing information, $2$ partial specification, $3$ mostly specified components, and $4$ sufficient detail for independent reconstruction.
\begin{equation}
    r(\cdot)\in\{1,2,3,4\}.
\end{equation}


The micro-level reproducibility score is determined with the LLM-based agentic extraction pipeline, distinguishing stage-specific aspects, where for example algorithm clarity or definition of hyper parameters play a role in method nodes but not in source nodes.
Extracting a set of stage-specific attributes (Appendix~\ref{appendix:ex_workflow}), providing a source quote from the paper with supporting evidence, and requiring a reasoning for the assessment, guide the LLM-based reproducibility assessment of nodes in a structured way.

\subsection{Reproducibility Scoring}

Given the inherent graph-structure of a scientific workflow inside a scientific paper, we propose to aggregate the micro-level assessments further into one final, more representative reproducibility score $R$, combing per-profile reproducibility indices along two orthogonal axes.
\textbf{The content score} $R_c \in [0,1]$ measures the descriptive sufficiency per-node, aggregating the analyst's rubric ratings on the sources, processes, and sink layers. 
\textbf{The structural score} $R_s \in [0,1]$ measures the quality of connectivity of the workflow graph extracted from the document, including the boundary connectivity of declared sources and sinks, the resolution of internal references, and the graph. 
The two axes are orthogonal, as $R_c$ captures whether each node is described in enough detail to be re-executed, while $R_s$ reflects whether the workflow is wired.
Their combined composite $R$ is their geometric mean, chosen so that weakness in either axis cannot be compensated by strength in the other:

\begin{equation}
    \text{R} = \sqrt{R_c \cdot R_s}
\end{equation}

\paragraph{Content Score.}         

The content score $R_c\in[0,1]$ measures whether the nodes of the workflow graph are described in sufficient detail for independent reconstruction.
Given the micro-level rubric score $r_{100}(v)\in[0,100]$ for each node $v\in V$, we first compute average reproducibility scores for each workflow stage $k\in\mathcal{K}$ that is present in the graph, rescaled to $[0,1]$, and then determine $R_c$ through a weighted convex combination over the present stages.                                                                                             
\begin{equation}       
    \mathcal{K}=\{\text{sources},\text{methods},\text{experiments},\text{sinks}\},                         \qquad                                        \mathcal{K}_{\text{present}} = \{k\in\mathcal{K} : V_k \neq \emptyset\}     
\end{equation}                                      
{\renewcommand{\arraystretch}{0}%
    \begin{table}[!h]                            \centering                                   \begin{tabularx}{1.014\textwidth}{XX}
    
        \begin{equation}
            R_k                                  =                                    \frac{1}{100\,|V_k|}                 \sum_{v\in V_k} r_{100}(v) 
        \end{equation} 
        &
        \begin{equation}
            R_c                                  =
            \sum_{k\in \mathcal{K}_{\text{present}}}
            \frac{w_k}{\sum_{j\in \mathcal{K}_{\text{present}}} w_j}   \, R_k
        \end{equation}                            
    \end{tabularx}                             
    \end{table}
}

Renormalization by $\sum_{j\in \mathcal{K}} w_j$ ensures that $R_c$ remains bounded in $[0,1]$ when one or more stages are absent.
This formulation captures the descriptive completeness of workflow components independently of their connectivity structure and therefore complements the structural score $R_s$, which evaluates dependency resolution and graph topology.
In this work, without loss of generality, we chose to weight these components with following fixed weights, where \textit{process} nodes comprise \textit{methods} and \textit{experiments}: $(w_{\text{sources}},\,w_{\text{process}},\,w_{\text{sinks}}) = (0.30,\,0.40,\,0.30)$.

\paragraph{Structural Score.} 
The structural score $R_s$ is the convex, weighted combination of five components as shown in Equation \eqref{eq:structural}.
Each component measures one structural failure mode of the extracted workflow.    
Essentially, the components from Equations~\eqref{eq:rsrc}--\eqref{eq:rlwcc} penalize disconnected sources, orphan sinks, unresolved references, broken source-to-sink paths, and further disconnected fragments.

\begin{equation}                                              
  R_s \;=\; w_{\mathrm{c1}}\,r_{\mathrm{c1}} + w_{\mathrm{c2}}\,r_{\mathrm{c2}} + w_{\mathrm{c3}}\,r_{\mathrm{c3}} + w_{\mathrm{c4}}\,r_{\mathrm{c4}} + w_{\mathrm{c5}}\,r_{\mathrm{c5}}
  \label{eq:structural}                             
\end{equation}


Component $r_{c1}$ (Equation~\ref{eq:rsrc}) measures the fraction of declared sources that are consumed by at least one process, penalizing unread inputs, where $deg^{out}(n)$ denotes the out-degree of node $n$. 
Component $r_{c2}$ (Equation~\ref{eq:rsink}) measures the fraction of declared sinks (figures and tables) that are produced by at least one process, penalizing orphan outputs, where $deg^{in}(n)$ denotes the in-degree of node $n$.
Component $r_{c3}$ (Equation~\ref{eq:rin}) reports the share of resolved input references (edges) into methods and experiments whose identifiers resolve to a known node, with unresolved references treated as dangling inputs $n_{d-i}$, that denote labels which do not match any node. 
Component $r_{c4}$ (Equation~\ref{eq:rreach}) is the probability that a uniformly chosen source-sink pair $(s,t)\in V_{\text{sources}}\times V_{\text{sinks}}$ is joined by a directed path in $G$, penalizing profiles in which source nodes cannot be traced to a sink node, where $[s\rightsquigarrow t]$ denotes the existence of a directed path. 
Component $r_{c5}$ (Equation~\ref{eq:rlwcc}) quantifies global cohesion as the share of nodes contained in the largest weakly connected component, penalizing fragmentation into disconnected sub-workflows, where $\mathrm{LWCC}(G)$ denotes the largest weakly connected component of the workflow graph $G$.
In this work, we chose to weight these components, without loss of generality, with following fixed weights summing to one: $(w_{\mathrm{c1}},w_{\mathrm{c2}},w_{\mathrm{c3}},w_{\mathrm{c4}},w_{\mathrm{c5}}) = (0.25,0.25,0.20,0.15,0.15)$. 

{\renewcommand{\arraystretch}{0}%
\begin{table}[!h]
    \centering
    \begin{tabularx}{1.014\textwidth}{XX}

\begin{equation} \label{eq:rsrc}  
    r_{\mathrm{c1}} = \frac{\bigl|\{n\in V_{\text{sources}}:\deg^{out}(n)>0\}\bigr|}{|V_{\text{sources}}|} 
\end{equation} 

&

\begin{equation} \label{eq:rsink}
    r_{\mathrm{c2}} = \frac{\bigl|\{n\in V_{\text{sinks}}:\deg^{in}(n)>0\}\bigr|}{|V_{\text{sinks}}|} 
\end{equation}

    \end{tabularx}
\end{table}
}

\begin{equation} \label{eq:rin}
    r_{\mathrm{c3}} = 1 - \frac{n_{d-i}}{\sum_{n\in \mathcal{N} }\deg^{in}(n)}, \;\;\;\; \text{with} \; \mathcal{N}=V_{\text{methods}} \cup V_{\text{experiments}} 
\end{equation}  

\begin{equation} \label{eq:rreach}
    r_{\mathrm{c4}} = \frac{1}{|\mathcal{N}_1| |\mathcal{N}_2|}\sum_{s\in \mathcal{N}_1}\sum_{t\in \mathcal{N}_2}\mathbf{1}[s\rightsquigarrow t] , \;\;\;\; \text{with} \; \mathcal{N}_1=V_{\text{sources}}, \;\; \mathcal{N}_2=V_{\text{sinks}} 
\end{equation} 

\begin{equation}  \label{eq:rlwcc} 
    r_{\mathrm{c5}} = \frac{|\mathrm{LWCC}(G)|}{|V|} 
\end{equation}

\textit{Process} receives the largest content weight because it contains most reproducibility-critical details (algorithms, parameters, tools). \textit{Sources} and \textit{sinks}, the workflow’s in- and out-faces, split the remainder equally. Structurally, the two ratios that govern end-to-end reproducibility (use of cited datasets and production of claimed artifacts) receive the highest equal weight. Given established consumption, resolved inputs, the reachability of the source-to-sink and the largest connected component characterize the quality of the topology. We treat these weights as a transparent prior rather than a tuned hyperparameter. Appendix~\ref{app:sensitivity} shows that our conclusions are robust to this choice.

\subsection{ReScience C Dataset}

\href{https://rescience.github.io/}{ReScience C} (ISSN 2430-3658) is an open-access, peer-reviewed journal dedicated to the reproducibility of research through the publication of independently developed, open-source implementations (more than 213 articles since 2015). 
The review process is conducted by researchers who actively validate the reproducibility and reusability of each contribution. 
By design, all materials -- including discussions and metadata -- are openly available and version-controlled, supporting continuous verification and extension of published work. 
The \texttt{ReScience C} dataset serves as a benchmark to evaluate the capability of the \texttt{ARA} framework to domain-agnostically assess reproducibility of scientific papers from multiple different domains, which allows us to compare automated \texttt{ARA}-based evaluations with expert-verified reproduction outcomes and measure their alignment with human judgments of reproducibility.
Details on \textit{ReScience C} and comparable datasets can be found in Appendix~\ref{appendix:benchmarks}.


    



\section{Experiments}

One way to evaluate agentic reproducibility assessment systems (addressing \textbf{RQ2}) is to examine the consistency of their outputs across models, decoding settings, and repeated runs (Section~\ref{sec:consistency-analysis}).
Such consistency analyses assess the stability of extracted workflow graphs, including the number and structure of nodes, as well as the robustness of node-level micro-scores and composite reproducibility indices.
Because LLM-based agent pipelines involve probabilistic inference, these quantities may vary across different model architectures, sampling temperatures, and stochastic executions, making stability an important indicator of methodological reliability.

Another way 
(addressing \textbf{RQ1} and \textbf{RQ3})
is to compare automated judgments with human reproducibility assessments derived from replication studies or expert annotations (Section~\ref{sec:benchmark}).
Such comparisons provide an estimate of how closely document-level reasoning systems approximate expert evaluations, although they rely on the assumption that human assessments constitute a reliable reference standard.
\textit{Importantly, human reproducibility assessments often incorporate external information sources such as repositories, search engines, and author communication, whereas \texttt{ARA} operates strictly at the document level.}
Moreover, the human reference labels (used here) are derived from full replication efforts that involve implementing methods, executing experiments, and validating reported results over extended periods of time.
These processes can reveal practical reproducibility barriers that are not observable from the publication text alone, making direct comparisons between document-level assessments and execution-based replication outcomes inherently conservative for \texttt{ARA}.
As a result, the following experiments evaluate how far structured workflow reconstruction alone can approximate expert reproducibility judgments under this restricted information setting.

\subsection{Consistency of Workflow Reconstruction and Reproducibility [RQ2]} \label{sec:consistency-analysis}



We evaluate the consistency and stability of \texttt{ARA}'s workflow graph reconstruction and reproducibility scoring across different LLM architectures, decoding settings, and repeated runs.
To this end, we execute the full pipeline multiple times on a representative subset of \texttt{ReScience C} articles (Table~\ref{tab:rescience-sample}) while varying model choice (Table~\ref{tab:model-overview}) and sampling temperatures $T$ (Table~\ref{tab:consistency-failure-rate}).
Across runs, we measure consistency of extracted workflow graph structure as well as agreement in node-level micro-scores $r(v)$, stage-level scores $R_k$, and aggregated reproducibility indices $(R_c, R_s, R)$.

Table~\ref{tab:workflow-consistency-graph} summarizes the graph-structural consistency analysis (Appendix~\ref{appendix:consistency-analysis}), comprising graph-edit-distance (GED), and count variability of nodes and edges.
For each paper--model--temperature configuration, variability is first computed across ten repeated runs and then aggregated across conditions.
Reported values therefore reflect the average stability of workflow reconstruction under stochastic inference.

\begin{table}[!ht]
\centering
\caption{\textbf{Workflow Graph Consistency For Different Models ($T=0$)}}
\label{tab:workflow-consistency-graph}
\begin{tabular}{lccccccc}
\toprule
 & & \multicolumn{6}{c}{\textbf{Node and Edge Count Variability}} \\
\cmidrule(lr){3-8}
\textbf{LLM (n runs)} & \textbf{GED} & {$E$} & {$V$} & {$V_{\text{sources}}$} & {$V_{\text{methods}}$} & {$V_{\text{experiments}}$} & {$V_{\text{sinks}}$} \\
\midrule
gemini-2.5-flash (77)       & 0.76 & 15.36 & 5.29 & 0.45 & 3.92 & 2.31 & 0.45 \\
gemini-2.5-pro (91)         & 0.76 & 3.08  & 2.05 & 0.64 & 1.22 & 1.06 & 0.50 \\
gemini-3-flash-preview (94) & 0.48 & 2.69  & 1.10 & 0.24 & 0.52 & 0.72 & 0.48 \\
gemini-3.1-pro-preview (96) & 0.30 & 2.47  & 0.73 & 0.05 & 0.28 & 0.62 & 0.13 \\
gpt-4.1 (100)                & 0.39 & 2.81  & 1.31 & 0.34 & 0.69 & 0.63 & 0.36 \\
\midrule
qwen3-32b (60)              & 0.18 & 2.57  & 2.60 & 0.00 & 1.09 & 1.43 & 0.53 \\
qwen3-8b (66)               & 0.17 & 2.58  & 0.84 & 0.07 & 0.33 & 0.32 & 0.47 \\
\bottomrule
\end{tabular}
\end{table}

\begin{table}[!ht]
\centering
\caption{\textbf{Reproducibility Score Consistency For Different Models ($T=0$)}}
\label{tab:repro-score-consistency}
\begin{tabular}{lcccccccc}
\toprule
& \multicolumn{8}{c}{\textbf{Reproducibility Assessment Score Variability}} \\
\cmidrule(lr){2-9}
& \multicolumn{5}{c}{\textbf{Micro-Level Assessment}} 
& \multicolumn{3}{c}{\textbf{Composite}} \\
\cmidrule(lr){2-6} \cmidrule(lr){7-9}
\textbf{LLM (n runs)} 
& \textbf{All} 
& {$V_{\text{sources}}$} 
& {$V_{\text{methods}}$} 
& {$V_{\text{experiments}}$} 
& {$V_{\text{sinks}}$} 
& {$R_c$} 
& {$R_s$} 
& {$R$} \\
\midrule
gemini-2.5-flash (77)       & 0.05 & 0.04 & 0.06 & 0.08 & 0.03 & 0.03 & 0.14 & 0.06 \\
gemini-2.5-pro (91)         & 0.07 & 0.03 & 0.11 & 0.13 & 0.10 & 0.06 & 0.13 & 0.08 \\
gemini-3-flash-preview (94) & 0.05 & 0.11 & 0.07 & 0.04 & 0.03 & 0.05 & 0.09 & 0.05 \\
gemini-3.1-pro-preview (96) & 0.02 & 0.01 & 0.04 & 0.03 & 0.01 & 0.02 & 0.04 & 0.03 \\
gpt-4.1 (100)                & 0.04 & 0.02 & 0.06 & 0.11 & 0.02 & 0.04 & 0.08 & 0.05 \\
qwen3-32b (60)              & 0.02 & 0.00 & 0.02 & 0.04 & 0.00 & 0.01 & 0.02 & 0.01 \\
qwen3-8b (66)               & 0.02 & 0.03 & 0.02 & 0.05 & 0.04 & 0.02 & 0.09 & 0.05 \\
\bottomrule
\end{tabular}
\end{table}

Table~\ref{tab:repro-score-consistency} summarizes the consistency of the reproducibility assessments derived from the extracted workflow graphs.
For each paper--model--temperature configuration, we report standard deviation across repeated runs for the stage-specific node scores $R_k$, and the aggregate scores $R_c$, $R_s$, and $R$.
These within-configuration variability estimates are then averaged across papers and temperatures.
Lower values therefore indicate that the same model assigns more stable reproducibility assessments across repeated executions.

Overall, these findings indicate that workflow graph reconstruction and reproducibility assessments are not driven by oracle-like randomness or hallucination, but exhibit substantial reliability and stability across runs, supporting the robustness and practical quality of the proposed framework.
Given the consistency of workflow graphs (Tables~\ref{tab:workflow-consistency-graph},~\ref{tab:workflow-consistency-graph-avg-temperature}), reproducibility assessments (Tables~\ref{tab:repro-score-consistency},~\ref{tab:repro-score-consistency-avg-temperature}), failure rates (Table~\ref{tab:consistency-failure-rate}), run times and costs (Table~\ref{tab:model-overview}), we found model \textit{gemini-3.1-pro-preview} (at temperature 0) the best, most cost-effective performing model to proceed further investigation with.

\subsection{Agentic Reproducibility Assessments Come Close To Human Perspectives [RQ1 \& RQ3]} \label{sec:benchmark}

        


We benchmark \texttt{ARA} against two existing agentic reproducibility assessment systems, \textit{ReplicatorAgent}~\citep{nguyen2026replicatorbench} and \textit{ReproScreener}~\citep{bhaskar2024reproscreener}, across three datasets: \texttt{ReScience C}, \textit{ReproBench}~\citep{hu2025repro}, and \textit{GoldStandardDB}~\citep{bhaskar2024reproscreener}, as shown in Table~\ref{tab:benchmark-comparison}.
These datasets differ in domain coverage, annotation protocols, and reproducibility scoring schemes, providing complementary evaluation settings for document-level assessment methods (Table~\ref{tab:dataset_comparison}).

\begin{table}[!ht]
    \centering
    \caption{\textbf{Reproducibility Assessment Benchmark.}}
    \label{tab:benchmark-comparison}
    \begin{tabular}{lccc}
        \toprule
        \textbf{Dataset} & \multicolumn{3}{c}{\textbf{Reproducibility Assessment Method}} \\
        \cmidrule(lr){2-4}
        \;\;\;\;(Metric) & \textbf{ARA} & \textbf{ReplicatorAgent} & \textbf{ReproScreener} \\
        \midrule
        {ReScience C} (213 articles) & & & \\
        \;\;\;\;ACC [\%] & 60.98 (26.41) & -- & -- \\
        \;\;\;\;F1 [\%] & 12.49 (18.24) & -- & -- \\
        \;\;\;\;Score Distance & 0.99 (0.81) & -- & -- \\
        \;\;\;\;Abs. Score Distance & 1.05 (0.74) & -- & -- \\
        {ReproBench} (112 articles) & & & \\
        \;\;\;\;ACC [\%] & \textbf{60.71} (17.96) & 36.84 & -- \\
        \;\;\;\;F1 [\%] & 13.32 (18.66) & \textbf{22.67} & -- \\
        \;\;\;\;Score Distance & 0.67 (1.32) & \textbf{0.63} & -- \\
        \;\;\;\;Abs. Score Distance & 1.19 (0.88) & \textbf{0.98} & -- \\
        {GoldStandardDB} (50 articles) & & &  \\
        \;\;\;\;ACC [\%] & \textbf{61.68 (32.94)} & -- & 43.56 (20.12) \\
        \;\;\;\;F1 [\%] & \textbf{50.07 (33.98)} & -- & 36.66 (24.30) \\
        \bottomrule
    \end{tabular}
\end{table}

\begin{table}[!ht]
    \centering
    \caption{\textbf{Stage-Wise Reproducibility Assessment on ReScience-C.}}
    \label{tab:rescience-ara-scores}
    \begin{tabular}{lcccc}
        \toprule
        \textbf{Stage} & \textbf{ACC} & \textbf{F1} & \textbf{Score Distance} & \textbf{Abs. Score Distance} \\
        \midrule
        Sources 
        & 62.24 (15.11) 
        & 20.04 (11.58) 
        & -0.89 (1.37) 
        & 1.32 (0.96) \\
        Methods 
        & 58.89 (21.29) 
        & 13.99 (11.26) 
        & 1.11 (0.91) 
        & 1.22 (0.76) \\
        Experiments 
        & 56.37 (24.71) 
        & 10.21 (8.91) 
        & 1.30 (0.87) 
        & 1.37 (0.75) \\
        Sinks 
        & 63.79 (21.83) 
        & 16.86 (19.30) 
        & 0.49 (1.04) 
        & 0.92 (0.69) \\
        \midrule
        Overall 
        & 60.98 (26.41) 
        & 12.49 (18.24) 
        & 0.99 (0.81) 
        & 1.05 (0.74) \\
        \bottomrule
    \end{tabular}
\end{table}

Across datasets, \texttt{ARA} achieves consistent agreement with human assessments, with accuracies around $60\%$ despite operating strictly under a document-level evaluation setting.
Performance remains stable even on the cross-domain \texttt{ReScience C} corpus, which represents the most heterogeneous and challenging benchmark considered in this study.
Compared with previously reported results for \textit{ReplicatorAgent} and \textit{ReproScreener}, \texttt{ARA} achieves comparable or higher agreement with human annotations across most evaluation settings.
Although \textit{ReplicatorAgent} achieves stronger performance on \textit{ReproBench} in terms of F1 score and distance metrics, this comparison must be interpreted cautiously because \textit{ReplicatorAgent} has access to external information sources, including internet retrieval and human replication reports, whereas \texttt{ARA} relies exclusively on the publication text.

To better understand disagreement patterns between agentic and human assessments, we further analyze stage-level reproducibility scores $R_k$ on \texttt{ReScience C} (Table~\ref{tab:rescience-ara-scores}).
Across workflow stages, \texttt{ARA} produces slightly more optimistic reproducibility scores than human annotators, with a mean signed score distance close to one ordinal scale level.
Agreement is highest for sources and sinks, where reporting structures are typically explicit in the document.
In contrast, agreement is lower for methods and experiments, which require reconstructing implicit procedural dependencies that often become apparent only during full replication attempts.
Because the human reference labels are derived from execution-based reproduction studies rather than document-level inspection alone, they capture implementation-level barriers that are not observable from text alone.

Overall, these results demonstrate that structured workflow reconstruction enables agentic systems to approximate human reproducibility judgments with moderate agreement despite operating without access to external artifacts, replication packages, or web-based information sources.
This supports the feasibility of document-level reproducibility assessment as a scalable complement to expert peer review.

\subsection{When Humans and Agents Disagree}



\begin{figure}[!h]
    \centering
    \includegraphics[width=\linewidth]{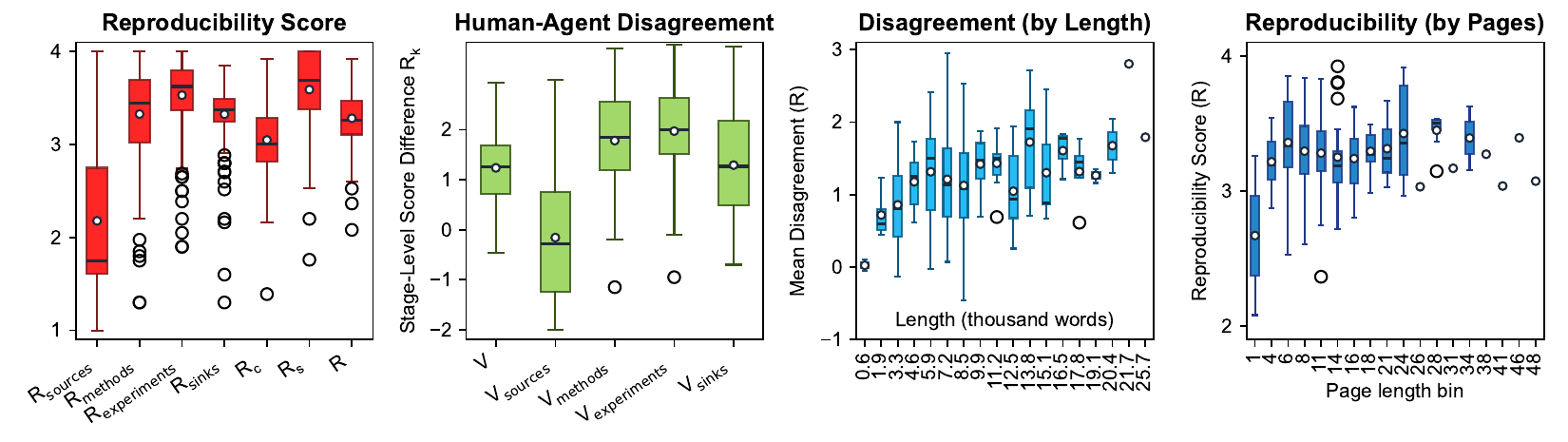}
    \caption{\textbf{Human-Agent Disagreement on Reproducibility Assessment (Rescience C).}}
    \label{fig:disagreement}
\end{figure}

To characterize where document-level workflow reconstruction diverges from execution-based human evaluations, we analyze signed score differences across workflow stages and document characteristics (Figure~\ref{fig:disagreement}).
Across the \texttt{ReScience~C} corpus, disagreement between \texttt{ARA} and human annotations remains bounded within approximately one ordinal scale level on average, which is expected given that \texttt{ARA} operates strictly on document-level evidence.
Agreement is highest for sources and sinks, where reporting structures are typically explicit, and lower for methods and experiments, which depend more strongly on implementation details revealed only during reproduction.
Short papers exhibit lower reproducibility scores overall, likely reflecting limited space for methodological specification, while disagreement is highest for medium-length papers and smaller for both very short and very long papers.
This pattern suggests that discrepancies arise primarily from differences in available procedural detail rather than instability of the workflow reconstruction procedure, supporting \texttt{ARA} as a scalable complement to execution-level reproducibility assessment.

\section{Discussion}

In this work, we demonstrated that workflow-based, document-level reproducibility assessments yield consistent scores, and approximate human judgments despite operating without access to external artifacts, repositories, or execution environments. 
The technical feasibility of state-of-the-art LLMs was validated in the experiments and benchmark. 
Therefore, agentic, document-level assessments could serve as a scalable diagnostic complement, not a replacement of human expert peer review.
This creates opportunities for the acceleration of editorial pipelines, large-scale literature screening systems, and meta-research infrastructure supporting transparency and reliability in scientific communication.
Acceptance of such assistance systems will, as we argue, largely depend on transparency; yet, as representative surveys show, LLMs are already used for peer-review.
Future work may extend the proposed framework beyond reproducibility assessment toward automated reproduction and implementation, integration of external artifact validation, and validity assessment of citations and scientific claims across publications.

    %
    
    
    %
    


\clearpage
\section*{Acknowledgements}

We gratefully acknowledge Jaime Suárez Corujo for kindly providing his paper as the seed case for our reproducibility analysis, and Ismini Dimitriadou for her assistance in securing GPT model credits.

Nikiforos Zacharof would like to specially thank Dimitrios Kourtesis from Ideas Forward for providing the tools and technical support that enabled the completion of this work. He also gratefully acknowledges Harry Kakoulidis for the insightful discussions and ideas that inspired and shaped the very early direction of this research.


We thank the Agentic Systems Lab (ASL) of ETH Zürich for supplying Gemini model credits.

We thank the Social Data Science Lab at the University of Konstanz for providing access to the GPUs used to run our local model analyses.

This work contributes to REproducible Research In Transportation Engineering (\href{https://www.rerite.org/}{RERITE}), advancing open science and transparency in transportation research, and the vision of a transportation community where open science, reproducible research and replicable studies are the norms, advancing scientific rigor, accelerating innovation and fostering translation into practice.
The authors acknowledge that discussions and community efforts within RERITE have contributed to shaping some of the ideas presented in this work. 
Michail A. Makridis also serves as a member of the RERITE steering committee.


\newpage
\bibliographystyle{IEEEtran}

\bibliography{references}

\clearpage
\appendix



\clearpage
\section{Appendix: Exemplary Scientific Workflow Graph} \label{appendix:ex_workflow}

Here, we provide an exemplary workflow generation for the work entitled \textit{"A Deep Reinforcement Learning Approach for Ramp Metering Based on Traffic Video Data"} from Bing et al. 2012 (doi: \href{https://doi.org/10.1155/2021/6669028}{10.1155/2021/6669028}) to illustrate the functioning of \texttt{ARA}.
This paper leverages real-world traffic data to generate a traffic simulation that is used to train a reinforcement-learning based controller, and its workflow graph is illustrated in the Figure~\ref{fig:example_workflow} below.

\texttt{ARA} distills source, method, experiment, and sink nodes, and their relationship (connection) from the document. 
Furthermore, the pipeline successfully identifies relevant sinks (actual results and outcomes of the research paper), while excluding irrelevant figures and tables (that rather serve for conceptual explanation or literature review purposes) from the analysis.

\begin{figure}[!h]
    \centering
    \includegraphics[width=\linewidth]{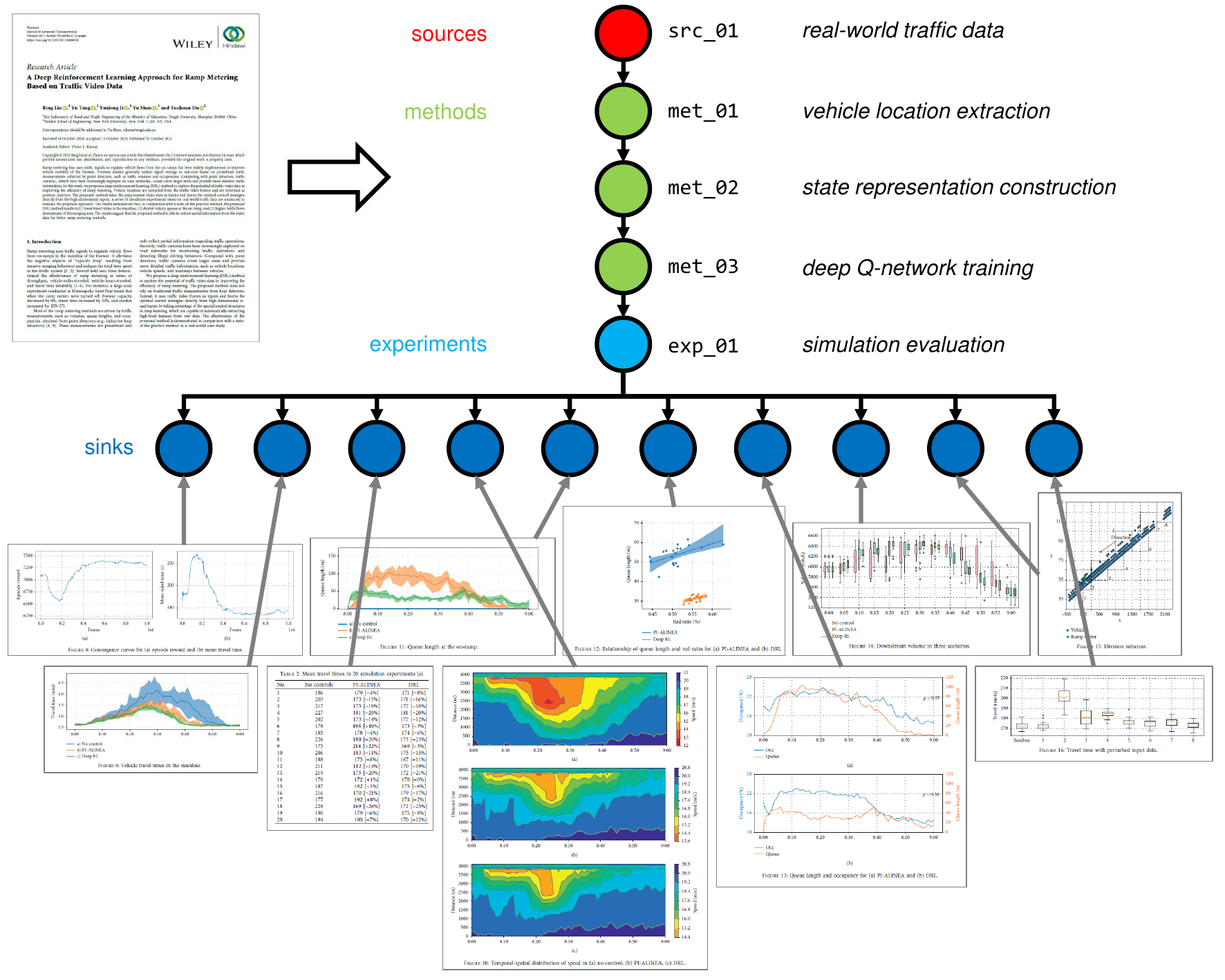}
    \caption{\textbf{Workflow Graph Generated From A Scientific Paper.}}
    \label{fig:example_workflow}
\end{figure}

The graph is stored in a Java-Script Notation Object (JSON) format as illustrated below. The general output structure comprises:
\begin{lstlisting}
{
  "metadata": { ... },
  "nodes_source": [ ... ],
  "nodes_process": [ ... ],
  "nodes_sink": [ ... ]
}
\end{lstlisting}

\newpage
The \textit{'metadata'} list comprises various general information about the publication, as well as a comprehensive list of \textit{'hyperparameters'}.
\begin{figure}[ht]
\noindent
\begin{minipage}[t]{0.6\textwidth}
\begin{lstlisting}[basicstyle=\ttfamily\small, frame=single]
"metadata": {
  "pdf_path": "...pdf",
  "extraction_model": "gemini-3.1-pro-preview",
  "title": "...",
  "authors": [ ... ],
  "repository_links": [ ... ],
  "supplementary_materials": [ ... ],
  "hyperparameters": [ ... ],
  "stated_software_versions": { ... },
  "hardware_requirements": null
},
\end{lstlisting}
\end{minipage}\hfill
\begin{minipage}[t]{0.39\textwidth}
\begin{lstlisting}[basicstyle=\ttfamily\small, frame=single]
"hyperparameters": [
  {
    "name": "...",
    "value": "...",
    "context": "...",
    "source_quote": "..."
  },
  ...
],
\end{lstlisting}
\end{minipage}
\end{figure}

\FloatBarrier

The \textit{'nodes\_process'} list comprises process nodes of type "method" and "experiment", while \textit{'nodes\_sink'} comprises nodes of type "figure" and "table".

\begin{figure}[!ht]
\noindent
\begin{minipage}[t]{0.49\textwidth}
\begin{lstlisting}[basicstyle=\ttfamily\small, frame=single]
"nodes_process": [
  {
    "node_id": "...",
    "node_name": "...",
    "source_quote": "...",
    "description": "...",
    "process_type": "method",
    "input_ids": [ ... ],
    "outcomes": [ ... ],
    "algorithm_clarity": 3,
    "tools_required": [ ... ],
    "tools_mentioned": [ ... ],
    "parameters_required": [ ... ],
    "parameters_mentioned": [ ... ],
    "reproducibility_score": 80,
    "reproducibility_rationale": "..."
  },
  ...
  {
    "node_id": "...",
    "node_name": "...",
    "source_quote": "...",
    "description": "...",
    "process_type": "experiment",
    "input_ids": [ ... ],
    "outcomes": [ ... ],
    "algorithm_clarity": 4,
    "tools_required": [ ... ],
    "tools_mentioned": [ .. ],
    "parameters_required": [ ... ],
    "parameters_mentioned": [ ... ],
    "reproducibility_score": 90,
    "reproducibility_rationale": "..."
  },
  ...
],
\end{lstlisting}
\end{minipage}\hfill
\begin{minipage}[t]{0.49\textwidth}
\begin{lstlisting}[basicstyle=\ttfamily\small, frame=single]
"nodes_sink": [
  {
    "node_id": "...",
    "node_name": "...",
    "source_quote": "...",
    "description": "...",
    "input_ids": [ ... ],
    "size_estimate": "...",
    "type": "figure",
    "statement_clarity": 3,
    "statement_validity": "...",
    "reproducibility_score": 85,
    "reproducibility_rationale": "..."
  },
  ...
  {
    "node_id": "...",
    "node_name": "...",
    "source_quote": "...",
    "description": "...",
    "input_ids": [ ... ],
    "size_estimate": "...",
    "type": "table",
    "statement_clarity": 4,
    "statement_validity": "...",
    "reproducibility_score": 85,
    "reproducibility_rationale": "..."
  },
  ...
],
\end{lstlisting}
\end{minipage}
\end{figure}

\clearpage
\section{Appendix: Prompt Templates} \label{appendix:prompts}

This appendix lists, verbatim, the prompts driving the structured extraction pipeline implemented in \texttt{src/ara\_pipeline/gemini\_rag.py}. The pipeline issues six queries per paper, all sharing one system instruction
(\ref{app:prompts:system}). Two of the six prompts are built at call time
(\ref{app:prompts:sink-nodes}, \ref{app:prompts:process-nodes}): the
canonical \texttt{src\_*} and \texttt{sink\_*} identifiers extracted by earlier queries are pasted into the prompt text so cross-layer wiring is resolved by exact string match. Execution order is: \texttt{header}
$\rightarrow$ \texttt{nodes\_source} $\rightarrow$ \texttt{nodes\_sink}
$\rightarrow$ \texttt{nodes\_process} $\rightarrow$ \texttt{artifacts}
$\rightarrow$ \texttt{parameters}.

\subsection{System Instruction (Shared by Every Query)}
\label{app:prompts:system}

\begin{lstlisting}
You are analysing a research paper for reproducibility. Answer using ONLY information explicitly stated in the attached PDF. If a fact is not stated, use null / omit it -- do NOT infer, guess, or draw on outside knowledge. Every item you extract must include a short literal quote from the paper that grounds it. Respond with valid JSON conforming to the provided schema; return nothing else.

BE TERSE. Avoid verbosity at all costs: do not restate the schema, do not add commentary, do not pad strings. Every string field is plain ASCII -- no combining diacritics, no non-BMP symbols, no long runs of repeated characters or tokens. Quotes are <=200 characters, descriptions <=400 characters, list items <=60 characters, lists <=8 items. If a field would exceed these limits, trim it; never produce filler to reach a limit. Close the JSON as soon as the required fields are populated -- truncation is a parse error.
\end{lstlisting}

\subsection{Header Prompt (Metadata)}
\label{app:prompts:header}

\begin{lstlisting}
Extract the paper's title and author list. If either is not clearly printed on the first page, use an empty string or empty list.
\end{lstlisting}

\subsection{Source Nodes Prompt (Datasets)}
\label{app:prompts:source-nodes}

\begin{lstlisting}
List every dataset referenced in this paper. For each dataset, emit an object with these fields (in this order):
  * node_id                    -- canonical 'src_<slug>' id (rules below)
  * node_name                  -- human-readable dataset name as printed in the paper
  * source_quote               -- short literal quote (<=200 chars) anchoring the dataset
  * description                -- description of contents (<=400 chars)
  * size_estimate              -- approximate size (string) if stated, else null
  * license                    -- licence string if stated, else null
  * availability               -- 'open', 'upon_request', or 'not_mentioned'
  * url                        -- URL or DOI if provided, else null
  * reproducibility_score      -- INTEGER percentage in [0, 100] (rubric below)
  * reproducibility_rationale  -- short justification (<=200 chars) for that score

REPRODUCIBILITY SCORE RUBRIC (critical):
- 0   -- the dataset is referenced but is NOT available: no URL, no DOI, no repository, no acquisition path stated; or the paper marks it as 'not_mentioned'.
- 100 -- the dataset is openly mentioned with a concrete, working source (URL / DOI / public repository) that a third party can follow to obtain the exact dataset.
- Intermediate values reflect partial information, e.g.:
    * ~25  -- only a citation to a prior paper, no URL or repository;
    * ~50  -- access 'upon request' from authors, no public link;
    * ~75  -- public repository named (e.g. 'available on Zenodo') but no direct URL / DOI given.
Always set 'reproducibility_rationale' to a short string citing the specific facts that drove the score (e.g. 'open URL provided', 'citation-only, no DOI', 'upon-request, no link'). Anchor the score in what is actually stated in the paper -- do not infer availability from outside knowledge.

NODE ID RULES (critical):
- 'node_id' is used verbatim by nodes_process[*].input_ids to wire sources into the workflow graph, so format and stability matter.
- Format: 'src_<slug>' where <slug> is lowercase ASCII, digits and underscores only, max 40 chars, derived from the most specific noun phrase naming this dataset in the paper (NOT from description or license). Examples: 'src_adni_t1', 'src_synthetic_reward_traces', 'src_mnist'. Do not include the year, version, or author name unless the paper uses it as part of the dataset's primary name.
- node_id MUST be unique across the returned list.
- Keep node_id stable: for the same dataset in the same paper, the same slug should be emitted every run.
\end{lstlisting}

\subsection{Sink Nodes Prompt (Results-Bearing Figures and Tables)}
\label{app:prompts:sink-nodes}

This prompt is built at call time by \texttt{build\_sink\_nodes\_prompt(source\_ids)};
the \texttt{SOURCE IDS} block is filled with the \texttt{src\_*} identifiers
returned by the source-nodes query.

\begin{lstlisting}
List ONLY the figures and tables that REPORT RESULTS of the study. Emit a sink_node ONLY for an artifact that presents an outcome, measurement, evaluation, comparison, or finding produced BY the paper's methodology -- the kind of artifact a replication would have to recompute to claim the result was reproduced.

EXCLUDE (do NOT emit a sink_node for these):
  * methodology / architecture diagrams, schematics, flowcharts, pipelines, conceptual figures, illustrative cartoons;
  * tables that list hyperparameters, configuration values, network architecture, dataset statistics, parameter ranges, software versions, hardware specifications, symbol/notation glossaries, or experimental setup;
  * tables of related work, summaries of prior literature, or comparisons of methods that predate the paper's contribution;
  * sample / qualitative example figures that show inputs rather than outcomes; legend or colour-key panels.
If unsure whether an artifact is a result, ask: 'Does this report a MEASURED OUTCOME of the study?' If no, EXCLUDE it. Information from excluded tables/figures (parameter values, methodology details) is captured downstream by the nodes_process extraction; do NOT duplicate it here.

For each remaining results-bearing figure or table, emit a single sink_node object with these fields:
  * node_id            -- canonical 'sink_fig<N>' or 'sink_tab<N>' (rules below)
  * node_name          -- the paper's printed label, e.g. 'Figure 3', 'Table 1'
  * source_quote       -- short literal quote (<=200 chars) from the caption or prose
  * description        -- what the figure/table shows or presents (<=400 chars)
  * input_ids          -- predecessor labels (rules below)
  * size_estimate      -- approximate size if stated (e.g. 'N=120 cells x 6 columns', '8 panels'), else null
  * type               -- 'figure' or 'table'
  * statement_clarity  -- ordinal reconstructability score r(.) on a 4-level scale (BARE INTEGER in {1, 2, 3, 4}):
      1 -- MISSING information: only the artifact is shown; no procedure or inputs are described.
      2 -- PARTIAL: the procedure is named with some detail, but at least one required input or step is absent or only loosely described.
      3 -- MOSTLY SPECIFIED: most steps and inputs are stated; minor gaps remain (e.g. an unstated default, an ambiguous parameter).
      4 -- SUFFICIENT detail for independent reconstruction: every step FROM 'input_ids' TO this artifact is stated precisely enough to reimplement without guessing.
  * statement_validity -- one of:
      'supported'            -- claims directly justified by the listed inputs;
      'partially_supported'  -- some claims justified, others not;
      'unsupported'          -- conclusions exceed what the inputs can support;
      'not_assessed'         -- insufficient information to judge.
  * reproducibility_score      -- INTEGER percentage in [0, 100] (rubric below)
  * reproducibility_rationale  -- short justification (<=200 chars) for that score

REPRODUCIBILITY SCORE RUBRIC (critical):
The score combines TWO factors:
  (a) UPSTREAM REPRODUCIBILITY -- how openly stated, traceable, and parameterised the sources and method steps feeding 'input_ids' are (raw datasets cited with URL/DOI? methods specified clearly? parameter values stated? tools named?). Treat this as the SUM / AGGREGATE of the upstream nodes that produce the inputs to this artifact: weak inputs cap the achievable score.
  (b) DESCRIPTION COHERENCE -- how well the figure/table's stated result follows from those inputs. A well-described result that cleanly aggregates / visualises its inputs scores high; a result that introduces unexplained data, missing intermediate steps, or claims beyond what the inputs can yield scores low.
Anchors:
  * 0   -- inputs are unavailable / unspecified, OR the artifact's claim does not follow from them.
  * 100 -- every input is fully reproducible (open data, fully specified methods) AND the claim follows directly and coherently from those inputs.
  * Intermediate values reflect partial information, e.g.:
      ~25 -- some inputs cited but not retrievable; coherence weak;
      ~50 -- about half of the upstream chain is reproducible and the result is plausibly derivable;
      ~75 -- inputs mostly reproducible and the result follows clearly, with only minor unstated steps.
Treat the score as a GEOMETRIC integration of (a) and (b) -- if either is near zero (unavailable inputs OR incoherent claim) the score should be near zero. Always set 'reproducibility_rationale' to a short string citing the specific upstream gaps and any input-vs-claim coherence issues that drove the value (e.g. 'inputs cited only, coherent claim', 'inputs open + URL, claim aggregates them directly', 'one input missing, claim partially supported').

NODE ID RULES (critical):
- Each figure: 'sink_fig<N>' (e.g. 'sink_fig3', 'sink_fig3a' for sub-panels).
- Each table:  'sink_tab<N>' (e.g. 'sink_tab1').
- node_id MUST be unique across the returned list.

INPUT ID RULES (critical):
- 'input_ids' lists the immediate predecessor artifacts the figure or table is built from (lists <=8 items, each <=60 characters).
- Use the source ids below VERBATIM whenever the artifact draws directly on a raw dataset.
- Otherwise use a short lowercase snake_case label that a method step would naturally emit (e.g. 'metrics_<dataset>', 'predictions_<dataset>', 'features_<dataset>'). Downstream method extraction will reuse these labels.

SOURCE IDS -- use these verbatim in 'input_ids' whenever the figure/table consumes a raw dataset:
  - <src_id_1>
  - <src_id_2>
  - ...

OUTPUT-SIZE LIMITS (hard):
- All strings plain ASCII; no combining diacritics or non-BMP symbols.
- 'source_quote' <=200 chars; 'description' <=400 chars; 'node_name' <=8 words.
\end{lstlisting}

\subsection{Process Nodes Prompt (Workflow Steps)}
\label{app:prompts:process-nodes}

This prompt is built at call time by
\texttt{build\_process\_nodes\_prompt(source\_ids, sink\_ids)}; the
\texttt{SOURCE IDS} and \texttt{SINK IDS} blocks are filled with the
identifiers returned by the source- and sink-nodes queries.

\begin{lstlisting}
List every step of the research workflow in sequential execution order. Each step is a 'process node'; classify it as either a method (algorithmic / computational procedure) or an experiment (controlled trial, evaluation, ablation, comparison, sweep). Assign each step a canonical 'node_id' of the form 'meth_<NN>' with a zero-padded two-digit ordinal, starting at 'meth_01' and incrementing by 1 for each subsequent step. Also provide a short 'node_name' (<=8 words) and a 'description' of what the step does. Include a short literal quote (source_quote) anchoring each step.

OUTPUT-SIZE LIMITS (hard):
- Every field is plain ASCII. Do NOT emit combining diacritics, non-BMP symbols, or long runs of repeated characters.
- 'source_quote': <=200 characters, a single literal sentence fragment copied verbatim from the PDF.
- 'description': <=400 characters.
- 'node_name': <=8 words.
- Each list ('input_ids', 'outcomes', 'tools_required', 'tools_mentioned', 'parameters_required', 'parameters_mentioned') has at most 8 items; each item is <=60 characters.
- Never repeat the same token more than twice in a row. If you are near a field limit, stop and close the JSON -- truncation is a parse error.

FIELD SEMANTICS:
- 'process_type': one of
    * 'method'     -- an algorithmic or computational procedure (e.g. preprocessing, feature extraction, model training, inference, optimisation, simulation step);
    * 'experiment' -- a controlled trial that produces measured outcomes (e.g. evaluation on a held-out set, ablation study, hyperparameter sweep, baseline comparison, statistical test).
- 'input_ids': labels consumed by this step (source ids, sink ids, or verbatim labels emitted by earlier steps' 'outcomes').
- 'outcomes': labels produced by this step. Most are INTERMEDIATE artifacts consumed by a later step (partitions, preprocessed data, features, trained models, prediction tensors, metrics, etc.). Only the FINAL step(s) in a chain emit a sink id. A step may emit multiple outcomes: any number of intermediates plus at most one sink id.
- 'tools_required': the GENERIC technique / capability needed to execute the step (e.g. 'gradient-boosted classifier', 'stratified k-fold cross-validation', '2-D convolutional encoder'). State these even if the paper does not name a concrete library.
- 'tools_mentioned': CONCRETE tools / libraries / frameworks the paper explicitly names for this step, with version numbers when stated (append 'version not stated' otherwise). Leave empty if the paper does not name a tool for this step.
- 'parameters_required': names of parameters the technique NEEDS to be reproducible (e.g. 'learning_rate', 'batch_size', 'random_seed', 'n_estimators'), regardless of whether the paper provides a value. Use lowercase snake_case parameter names.
- 'parameters_mentioned': parameters the paper actually states for this step, as 'name=value' strings (e.g. 'learning_rate=1e-3', 'n_folds=5'). Only include parameters whose value is explicitly stated in the paper.
  IMPORTANT: tables and figures that are NOT results (e.g. tables of hyperparameters, configuration tables, architecture diagrams, notation glossaries, dataset-statistics tables, methodology flowcharts) are NOT nodes_sink -- they are part of the methodology specification. MINE them: extract every parameter value listed in such tables into 'parameters_mentioned' of the step that uses the parameter, fold any algorithmic detail shown in such figures into the step's 'description', and add any tool / library named there to 'tools_mentioned'. Anchor each extracted fact with a 'source_quote' from the table caption, table cell, or figure caption.
- 'algorithm_clarity': ordinal reconstructability score r(.) on a 4-level scale (integer in {1, 2, 3, 4}):
    * 1 -- MISSING information: the paper names the step but provides no algorithmic detail, no inputs, and no parameter values.
    * 2 -- PARTIAL specification: the algorithm is named with some detail, but at least one required input or parameter is absent or only loosely described.
    * 3 -- MOSTLY SPECIFIED components: most algorithmic choices, inputs, and required parameters are stated; minor gaps remain (e.g. an unstated default, an ambiguous tie-break).
    * 4 -- SUFFICIENT detail for independent reconstruction: the step's algorithm, inputs, and all required parameters are stated precisely enough to reimplement without guessing.
  Emit the BARE INTEGER (1, 2, 3, or 4) -- not a string.
- 'reproducibility_score': INTEGER percentage in [0, 100] derived from the four fields above, per this rubric:
    * 0   -- 'algorithm_clarity'=1 AND no tools / parameters listed.
    * 100 -- 'algorithm_clarity'=4 AND every item in 'tools_required' also appears in 'tools_mentioned' AND every item in 'parameters_required' has a value in 'parameters_mentioned'.
    * Intermediate values reflect partial specification. Suggested anchors:
        ~25 -- 'algorithm_clarity'=2 with most tools / parameters missing;
        ~50 -- 'algorithm_clarity'=3 with about half the tools / parameters named;
        ~75 -- 'algorithm_clarity'=3-4 with most tools named and most parameter values stated, only minor gaps remain.
  Treat the score as the GEOMETRIC integration of (clarity / 4), (tools_named_ratio), and (params_valued_ratio) -- if any of the three is zero, the score should be near zero. Do not round to a flat 50 when the inputs disagree.
- 'reproducibility_rationale': short string (<=200 chars) citing the concrete gaps that drove the score, e.g. 'clarity=3, 1/2 tools named, 4/5 params stated' or 'clarity=4, all tools+params stated'.

SOURCE IDS -- use these verbatim in 'input_ids' whenever a step consumes a raw dataset. Do NOT use human-readable dataset names, and do NOT invent new 'src_*' ids:
  - <src_id_1>
  - <src_id_2>
  - ...

SINK IDS -- use these verbatim in 'outcomes' ONLY when the step's direct output IS one of the figures/tables reported in the paper (typically the last step in a chain). If the step produces an intermediate artifact that a later step will turn into a figure or table, DO NOT put a sink id on this step. Do NOT invent new 'sink_*' ids:
  - <sink_id_1>
  - <sink_id_2>
  - ...

OUTCOME NAMING TEMPLATES (critical):
When a step derives an artifact from a source dataset, name the outcome using these templates, where <slug> is the source node_id WITH THE 'src_' PREFIX REMOVED (e.g. for source 'src_adni_t1' use slug 'adni_t1'):
  * 'train_<slug>', 'val_<slug>', 'test_<slug>' for splits
  * 'preprocessed_<slug>' for cleaned / normalised data
  * 'features_<slug>' for extracted features
  * 'predictions_<slug>' for model outputs evaluated on that dataset
  * 'metrics_<slug>' for evaluation metrics computed over that dataset
If the paper uses a different but specific name for the artifact, you MAY use it; otherwise apply these templates so downstream steps can reference the artifact by a predictable name. Purely internal intermediates (not tied to any one source dataset) may use free-form lowercase snake_case names.

LABEL CONSISTENCY RULES (critical):
- Use the SAME label for the same artifact across steps. If step N produces an outcome that feeds step N+1, the string in step N's 'outcomes' MUST appear verbatim in step N+1's 'input_ids'. Do not paraphrase, pluralise, or reorder words between steps; pick one canonical name per artifact and reuse it.
- When a step splits, partitions, or divides a dataset (e.g. train/test split, k-fold, train/val/test, stratified sampling), 'outcomes' MUST list each resulting partition as a separate named item using the 'train_<slug>' / 'val_<slug>' / 'test_<slug>' templates above (use whatever names the paper uses if stated). The 'description' MUST state the split percentages / proportions / fold count exactly as reported in the paper (e.g. '80/20 split', '70/15/15', '5-fold cross-validation'). Downstream steps that consume a partition MUST reference it by the same name in 'input_ids'.
\end{lstlisting}

\subsection{Artifacts Prompt (Code, Supplements, Hardware, Metadata)}
\label{app:prompts:artifacts}

\begin{lstlisting}
List all code repositories, supplementary materials, and external resources mentioned anywhere in this paper -- references, footnotes, acknowledgements, data availability statement. Include URLs, GitHub links, Zenodo DOIs, and institutional repository references. Also state any stated hardware requirements.
\end{lstlisting}

\subsection{Parameters Prompt (Hyperparameters, Software, Hardware, Metadata)}
\label{app:prompts:parameters}

\begin{lstlisting}
List all parameters, hyperparameters, configuration values, and hardware requirements mentioned in this paper. Include numerical values, ranges, and any stated defaults. Return software versions as a list of objects, each with 'name' and 'version' fields.
\end{lstlisting}

\clearpage

\clearpage
\section{Sensitivity of Reproducibility Index to Weight Choice}                                                     \label{app:sensitivity}



The composite score $R = \sqrt{R_c \cdot R_s}$ depends on two convex weight vectors that we fix a priori rather than learn: the layer weights $w_j = \left(0.30, 0.40, 0.30\right)$ over (\texttt{source}, \texttt{process}, \texttt{sink}), which enter the content score $R_c$ with the process layer covering both methods and experiments and the structural weights $w_{ci} = \left(0.25, 0.25, 0.20, 0.15, 0.15\right)$ over (\texttt{sources\_consumed}, \texttt{sinks\_produced}, \texttt{resolved\_inputs}, \texttt{source-to-sink reachability}, \texttt{lwcc}), which enter the structural score $R_s$. 

Because these weights are design choices, a natural concern is whether \texttt{ARA}'s conclusions are an artifact of this particular prior. We therefore perform a sensitivity analysis: we perturb the weights along three independent axes and ask how much the resulting scores and, more importantly, the ranking of papers change. The analysis is run over all $N = 2{,}737$ profiles produced by the consistency study of Sec.~\ref{sec:consistency-analysis} on the ReScience,C benchmark, i.e.\ every (paper, analyst, repetition) triple obtained by sweeping seven analyst backends (Gemini 2.5 Flash/Pro, Gemini 3 Flash, Gemini 3.1 Pro, GPT-4.1, and the offline models Qwen3-8B and Qwen3-32B) at $T = 0$ with ten repetitions per paper. The three perturbation axes are:

\begin{itemize}  
    \item \textbf{Uniformization:} replace each weight vector with the uniform distribution; the most extreme symmetric prior, and a sanity check that no single weight dominates the result.
    
    \item \textbf{Leave-one-out (LOO):} zero one component at a time and renormalise; isolates the marginal contribution of each individual weight.                                                       
    \item \textbf{Dirichlet sampling:} draw random convex weights concentrated around the default; characterises the score distribution under continuous perturbation rather than the discrete schemes above. 
\end{itemize}

For each perturbation we report two complementary metrics: the per-profile score deviation $|\Delta R|$ against the default, which measures how much an individual paper's score shifts, and the per-paper Spearman rank correlation $\rho$ between the default and perturbed rankings, which measures whether the relative ordering of papers (the primary use case of \texttt{ARA}) is preserved. 

\paragraph{Discrete Schemes (Table~\ref{tab:weight-discrete}).}                  
Replacing both weight vectors with a uniform prior shifts the per-profile reproducibility index by only $|\Delta R| = 0.014$ on average ($p_{95} = 0.037$, max $= 0.058$), well below the within-configuration LLM-decoding variability already reported in Table~\ref{tab:repro-score-consistency}. Uniformizing either vector individually is even smaller ($|\Delta R| \leq 0.009$). Leave-one-out shifts are larger by construction, but identify the load-bearing components: removing the \emph{process} layer produces the largest mean shift ($|\Delta R| = 0.056$), while removing the \emph{sources\_consumed} ratio produces the largest mean structural shift ($|\Delta R| = 0.043$).

\begin{table}[!ht]                              
\centering                  
\caption{\textbf{Per-Profile Score Deviation Under Reweighting.}
  (mean / p95 / max of $|R - R_\text{default}|$ across $N{=}3{,}147$ profiles.)}     
  \label{tab:weight-discrete}                 
  \small                                  
  \begin{tabular}{lcccc}                   
  \toprule                 
  \textbf{Scheme} & $\bar{R}$ & mean $|\Delta R|$ & $p_{95} |\Delta R|$ & max $|\Delta R|$ \\                    
  \midrule                     
  default                              & 0.696 & 0.000 & 0.000 & 0.000 \\               
  uniform layer ($w_{ci} = 1/3$)   & 0.691 & 0.009 & 0.023 & 0.034 \\          
  uniform structural ($w_{s} = 1/5$) & 0.692 & 0.007 & 0.028 & 0.049 \\                 
  \textbf{uniform both}                & 0.688 & \textbf{0.014} & \textbf{0.037} & \textbf{0.058} \\
  \midrule          
  \multicolumn{5}{l}{\emph{Layer leave-one-out}} \\       
  ~~drop source                        & 0.729 & 0.047 & 0.137 & 0.158 \\               
  ~~drop process (method or experiment)                      & 0.664 & \textbf{0.056} & 0.154 & 0.524 \\                                   
  ~~drop sink                          & 0.683 & 0.026 & 0.071 & 0.281 \\                           
  \midrule                        
  \multicolumn{5}{l}{\emph{Structural leave-one-out}} \\
  ~~drop sources\_consumed             & 0.719 & 0.043 & 0.112 & 0.177 \\                    
  ~~drop sinks\_produced               & 0.680 & 0.035 & 0.089 & 0.236 \\                         
  ~~drop resolved\_inputs              & 0.663 & 0.036 & 0.105 & 0.232 \\                             
  ~~drop reachability                  & 0.727 & 0.031 & 0.061 & 0.066 \\                    
  ~~drop lwcc                          & 0.681 & 0.021 & 0.048 & 0.274 \\                  
\bottomrule              
\end{tabular}               
\end{table}

\paragraph{Rank stability (Table~\ref{tab:weight-rank}).}   A more downstream-relevant test is whether the \emph{cross-paper ordering} survives reweighting. The replacement of both vectors with a uniform prior preserves the rank in Spearman $\rho = 0.95$; either vector alone preserves it in $\rho \geq 0.98$. Leave-one-out additionally identifies which components carry the most ranking information: dropping \emph{sources\_consumed} reduces $\rho$ to $0.84$, the largest single-component impact in the table. The metric \emph{resolved\_inputs}, by contrast, leaves the ranking unchanged ($\rho = 1.00$), indicating partial redundancy with the remaining structural metrics on this corpus a target for future simplification of the index.

\begin{table}[!ht]                              
\centering              
\caption{\textbf{Spearman Rank Correlation Against the Default Ranking.}
  Per-paper mean $R$ aggregated across all (model, temperature, sample) triples.}  
  \label{tab:weight-rank}                     
  \small                                  
  \begin{tabular}{lc}                     
  \toprule                   
  \textbf{Scheme} & $\rho_\text{vs default}$ \\
  \midrule              
  uniform layer            & 0.98 \\           
  uniform structural       & 0.99 \\            
  \textbf{uniform both}    & \textbf{0.95} \\        
  \midrule                                        
  LOO layer:source         & 0.88 \\          
  LOO layer:process        & 0.89 \\              
  LOO layer:sink           & 0.90 \\                    
  \midrule                                    
  LOO struct:sources\_consumed   & \textbf{0.84} \\         
  LOO struct:sinks\_produced     & 0.98 \\        
  LOO struct:resolved\_inputs    & 1.00 \\            
  LOO struct:reachability        & 0.95 \\        
  LOO struct:lwcc                & 0.99 \\            
  \bottomrule                         
\end{tabular}
\end{table}

\begin{figure}[!h]
    \centering
    \includegraphics[width=\linewidth]{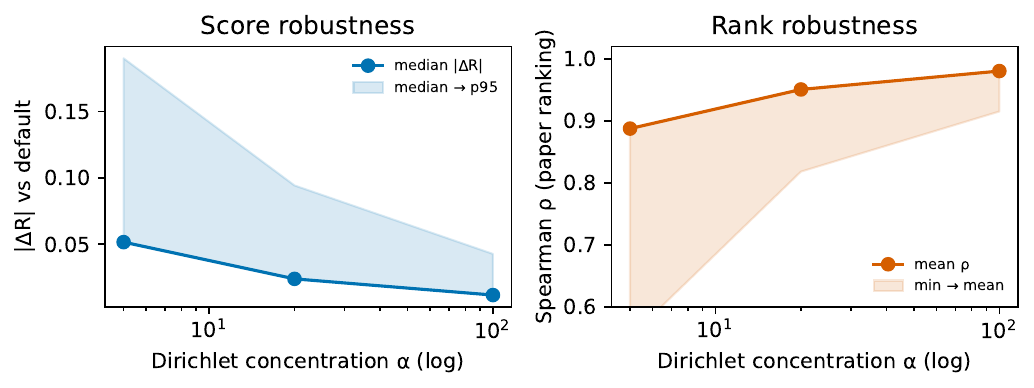}
    \caption{\textbf{Score and Rank Robustness under Dirichlet Perturbation.} Left: per-profile $|\Delta R|$ vs default (median, p95 band) as a function of concentration $\alpha$. Right: per-paper Spearman $\rho$ vs default (min, mean band) across $200$ draws per $\alpha$. Both axes are log-scaled in $\alpha$.}
    \label{fig:weight-dirichlet}
\end{figure}

\paragraph{Continuous Dirichlet Sweep (Table~\ref{tab:weight-dirichlet}, Figure~\ref{fig:weight-dirichlet}).}
We further sample $K = 200$ weight vectors per concentration $\alpha \in \{5, 20, 100\}$ from $\mathrm{Dir}(\alpha \cdot w_\text{default})$, so the expectation of every draw equals the default while the spread is controlled by $\alpha$. At realistic perturbation widths ($\alpha = 20$, equivalent to weight uncertainty $\approx \pm 0.05$), the median per-profile $|\Delta R| = 0.024$ and the worst-case Spearman $\rho$ across the $200$ draws is $0.82$. Restricting to $\alpha = 100$ recovers $\rho \geq 0.92$ for every draw. Even under the wide setting $\alpha = 5$, the \emph{mean} rank correlation remains $0.89$, establishing a lower bound on robustness against adversarial weight choices near the default.

\begin{table}[!ht]                              
  \centering
  \caption{\textbf{Dirichlet Perturbation Around the Default Weights.}
  $K{=}200$ draws per concentration; $|\Delta R|$ pooled across all draws and profiles, $\rho$ summarized across draws.}
  \label{tab:weight-dirichlet}                
  \small                                      
  \begin{tabular}{ccccc}             
  \toprule                     
  $\alpha$ & median $|\Delta R|$ & $p_{95} |\Delta R|$ & mean $\rho$ & min $\rho$ \\                     
  \midrule           
  5    & 0.052 & 0.190 & 0.89 & 0.55 \\
  20   & 0.024 & 0.094 & 0.95 & 0.82 \\       
  100  & 0.012 & 0.043 & 0.98 & 0.92 \\                         
  \bottomrule                                 
  \end{tabular}                                          \end{table}

The three sensitivity axes converge on the same conclusion: \texttt{ARA}'s output is governed by the structure of the workflow graph and the analyst's per-node scores, not by the specific values in $w_j$ or $w_{ci}$. Uniformization shifts the index by $|\Delta R| = 0.014$ on average (below the within-configuration decoding noise of Table~\ref{tab:repro-score-consistency}) and the cross-paper ranking survives every realistic perturbation with $\rho \geq 0.95$. We therefore keep the default weights as the headline configuration: they are interpretable, lie well within the stable weight region, and re-tuning them would change the score less than the analyst’s own stochasticity. Three directions remain. \emph{(i)} The LOO ranking analysis shows that \texttt{resolved\_inputs} ($\rho = 1.00$) is partially redundant with other structural metrics on this corpus, suggesting a leaner four-component $R_s$ for testing on other benchmarks. \emph{(ii)} In contrast, \texttt{sources\_consumed} ($\rho = 0.84$) and the \emph{process} layer contain the most ordering information and merit a more informative prior. \emph{(iii)} Once a large enough set of papers with human reproduction outcomes is available, the weights can be \emph{learned} (e.g.\ by maximising rank agreement with human verdicts) rather than fixed, turning the current sensitivity bound into a calibrated estimate.

\clearpage
\section{Appendix: Benchmarks \& Datasets} \label{appendix:benchmarks}

This subsection describes the benchmark datasets used to evaluate our reproducibility assessment framework. 
We rely on three complementary resources that span different scientific domains, annotation schemes, and levels of granularity: \textit{GoldStandardDB}~\citep{bhaskar2024reproscreener}, \textit{Repro-Bench}~\citep{hu2025repro}, and the \textit{ReScience~C} corpus~\citep{rougier2018rescience}. 
Together, these datasets provide manually curated ground-truth reproducibility annotations derived from expert assessments and reproduction reports, enabling evaluation across both binary and ordinal scoring settings. 
Table~\ref{tab:dataset_comparison} summarizes their key characteristics, and the following subsections describe each dataset in more detail.

Due to differences in scoring formats, artifact availability, and implementation constraints, it was not possible to evaluate \textit{ReplicatorAgent} and \textit{ReproScreener} on \texttt{ReScience C} in Table~\ref{tab:benchmark-comparison}.

\begin{table}[!h]
\renewcommand{\tabularxcolumn}[1]{>{\raggedright\arraybackslash}p{#1}}
\centering
\caption{\textbf{Reproducibility Assessment Benchmark Datasets.} This table summarizes and contrasts three major reproducibility assessment benchmark datasets used in this study.}
\label{tab:dataset_comparison}
\small
\begin{tabularx}{\linewidth}{lXXX}
\toprule
    \textbf{Criterion} & \textbf{GoldStandardDB}~\citep{bhaskar2024reproscreener} & \textbf{Repro-Bench}~\citep{hu2025repro} & \mbox{\textbf{ReScience C}~\citep{rougier2018rescience} (this work)} \\
    \midrule
    
    Size (\# papers)
    & 50
    & 112
    & 213
    \\
    
    Time-span
    & 2021--2022
    & 2019--2024
    & 2015--2026
    \\
    
    Domain
    & Machine Learning
    & Social Sciences
    & Multi-Domain
    \\
    
    Reproducibility
    \\
    
    \makecell[tl]{\;\;\;\; Scale}
    & Binary (0 / 1)
    & Ordinal (1--4)
    & Ordinal (1--4)
    \\
    
    \makecell[t]{\;\;\;\; Dimensions}
    & \makecell[tl]{Gunderson metrics~\citep{gundersen2018state}\\ objective, method,\\dataset, hypothesis, \\prediction, code, setup)}
    & \makecell[tl]{Total Score}
    & \makecell[tl]{Sources, Methods,\\Experiments, Sinks}
    \\
    
    Source
    & \makecell[tl]{arXiv preprints\\ (\texttt{cs.LG}, \texttt{stat.ML})}
    & \makecell[tl]{Mass reproduction study\\I4R discussion papers\\Retraction Watch database\\X (Twitter)}
    & \makecell[tl]{ReScience C open-\\ access replication journal}
    \\
\bottomrule
\end{tabularx}
\end{table}

\subsection{GoldStandardDB}

The \textit{GoldStandardDB} dataset~\citep{bhaskar2024reproscreener} consists of 50 machine learning preprints sampled from arXiv, specifically drawn from the \texttt{cs.LG} and \texttt{stat.ML} categories, which represent the principal domains for machine learning research submissions on the platform. 
The preprints were selected using a controlled query restricted to manuscripts last updated between October 24 and October 25, 2022, comprising 50 manuscripts. 
Each paper was manually labeled by the authors with respect to a predefined set of binary reproducibility indicators derived from the Gunderson metrics~\citep{gundersen2018state}, including statements of (i) research problem, (ii) research objective, (iii) research method, (iv) dataset, (v) hypothesis, (vi) predictions, (vii) code availability, and (viii) experiment setup.

\subsection{Repro-Bench}

The \textit{Repro-Bench} dataset~\citep{hu2025repro} is a curated benchmark consisting of 112 scientific works, each corresponding to a published social science paper accompanied with its publicly available (human generated) reproduction package and reproduction report. 
The dataset spans multiple subdomains of social science research and was constructed from four complementary sources: a \textit{large-scale mass reproduction effort}~\citep{brodeur2024mass} covering economics and political science papers (92 instances), the \textit{Institute for Replication (I4R) discussion paper series}~\citep{i4r} (11 instances), the\textit{ Retraction Watch database}~\citep{retrdb} (7 instances), and publicly documented reproduction analyses shared on \textit{X} (formerly \textit{Twitter}) (2 instances). 
Each paper satisfies strict inclusion criteria, including the availability of a DOI, an executable reproduction package (data, code, and documentation), and a credible expert-written reproduction report verifying computational results. 
Ground-truth reproducibility labels are derived directly from these reproduction reports and assigned using a four-level scoring scheme that reflects the consistency between reported findings and reproduced outputs: score~1 indicates irreproducible major findings, score~2 indicates minor coding inconsistencies without affecting conclusions, score~3 indicates minor reporting discrepancies (e.g., rounding differences), and score~4 indicates fully reproducible results. 
The annotation process follows a structured consensus-based protocol in which the lead author first assigns labels based on reproduction reports and a five-member expert team subsequently cross-validates all scores to ensure consistency~\citep{hu2025repro}. 

\subsection{ReScience C}

We additionally leverage the \textit{ReScience~C} corpus, a platinum open-access, peer-reviewed journal dedicated to the publication of reproducible replications of computational research \citep{rougier2018rescience}. 
ReScience~C specifically targets independent reimplementations of previously published studies and requires authors to provide complete open-source code, documentation, and replication reports verifying whether the original results can be reproduced. 
Unlike traditional journals, ReScience~C operates through an open and collaborative publishing workflow hosted on GitHub, where submissions, reviews, code execution checks, and revisions are conducted transparently and publicly. 
As of 2015–2026, the journal contains 213 published replication studies spanning multiple computational domains. 
Each article includes a human-generated reproduction report together with executable code and supporting materials, allowing direct verification of whether figures, tables, and claims from the target paper can be reproduced. 
These expert-written reproduction reports serve as ground-truth evidence for reproducibility assessments, making the corpus particularly suitable for benchmarking automated reproducibility evaluation methods.
Following \citep{hu2025repro} and \citep{brodeur2024mass}, we operationalize reproducibility as a structured, ordinal assessment reflecting the consistency between reported results and independently reproduced outputs. Specifically, we evaluate reproducibility on a four-level scale from 1 to 4 and measure reproducibility across four complementary dimensions that correspond to distinct stages of the evidence-generation pipeline (i) sources, (ii) methods, (iii) experiments, and (iv) sinks (results).
The labels are derived from the human generated reproducibility efforts, similar to~\citep{hu2025repro}.

\clearpage
\section{Appendix: Consistency Analysis} \label{appendix:consistency-analysis}

To evaluate the stability of workflow graph reconstruction and reproducibility scoring, we execute the full \texttt{ARA} pipeline repeatedly under controlled variations of model architecture, sampling temperature, and stochastic decoding.

\subsection{Document Selection}
Consistency experiments are performed on ten representative articles sampled from the \texttt{ReScience C} corpus as shown in Table~\ref{tab:rescience-sample}.
The selected documents span a broad range of lengths and methodological complexity to ensure coverage of heterogeneous workflow structures.

\begin{table}[!h]
    \centering
    \caption{\textbf{Representative ReScience C Articles Used for Consistency Analysis.}}
    \label{tab:rescience-sample}
    \begin{tabularx}{\linewidth}{lXcccc}
        \toprule
        \textbf{Paper} & \textbf{Domain} & \textbf{\#Pages} & \textbf{\#Words} & \textbf{\#Figures} & \textbf{\#Tables} \\
        \midrule
        2017\_04 \href{https://www.doi.org/10.1038/nature04605}{[DOI]} & evolutionary biology & 4  & 3,372  & 3  & 0 \\
        2020\_26 \href{https://www.doi.org/10.1016/j.tpb.2009.10.003}{[DOI]} & evolutionary biology & 8  & 3,701  & 2  & 0 \\
        2017\_01 \href{https://www.doi.org/10.1162/neco.2008.12-07-680}{[DOI]}  & computational neuroscience & 15 & 6,436  & 2  & 1 \\
        2025\_03 \href{https://www.doi.org/10.48550/arXiv.1905.11926}{[DOI]} & machine learning (vision) & 20 & 8,930  & 12 & 4 \\
        2017\_08 \href{https://www.doi.org/10.1016/j.neuroimage.2014.09.053}{[DOI]} & neuroimaging & 11 & 9,237  & 9  & 6 \\
        2022\_30 \href{https://www.doi.org/10.48550/arXiv.2011.00844}{[DOI]} & computer vision & 18 & 9,800  & 14 & 12 \\
        2021\_34 \href{https://www.doi.org/10.1111/1468-0262.00112}{[DOI]} & economics & 28 & 13,493 & 5  & 0 \\
        2023\_17 \href{https://openreview.net/forum?id=wbPObLm6ueA}{[URL]} & machine learning & 24 & 14,733 & 8  & 0 \\
        2023\_15 \href{https://www.doi.org/10.48550/arXiv.2203.01928}{[DOI]} & machine learning & 30 & 16,170 & 14 & 9 \\
        2022\_38 \href{https://openreview.net/forum?id=dlEJsyHGeaL}{[URL]} & graph machine learning & 34 & 22,917 & 11 & 4 \\
        \bottomrule
    \end{tabularx}
\end{table}

\subsection{Model Configurations}
We evaluate a selection of seven state-of-the-art LLMs, that include both hosted frontier systems and locally executed open-source, light-weight architectures, summarized in Table~\ref{tab:model-overview}.
Because local models have more limited context capacity than commercial systems, especially for documents exceeding approximately 15{,}000 words, some configurations exhibit reduced robustness on longer papers. To fit qwen3-32b on a single H100 GPU, its context window was reduced to 28K tokens, which further constrains its ability to process the longest papers in the benchmark.               
Including both local and hosted models therefore enables consistency comparisons across heterogeneous inference settings and supports selecting stable configurations for large-scale evaluation.

\begin{table}[!h]
    \centering
    \caption{\textbf{Language Models Used in Consistency Experiments.}}
    \label{tab:model-overview}
    \begin{tabular}{lccccc}
    \toprule
    & & & & \multicolumn{2}{c}{\textbf{Average (per doc.)}} \\
    \cmidrule(lr){5-6}
    \textbf{Model} & \textbf{Params} & \textbf{Context} & \textbf{Environment} & \textbf{Runtime} & \textbf{Cost} \\
    \midrule
    gemini-2.5-flash        & $\sim$5B (est.)  & $\sim$1M tokens  & hosted & 203.98 s & \$0.0767 \\
    gemini-2.5-pro          & n/a              & $\sim$2M tokens  & hosted & 180.10 s & \$0.3114 \\
    gemini-3-flash-prev.    & n/a              & $\sim$1M tokens  & hosted & 218.23 s & \$0.0767 \\
    gemini-3.1-pro-prev.    & n/a              & $\sim$2M tokens  & hosted & 365.20 s & \$0.3114 \\
    gpt-4.1                 & $\sim$12B (est.) & $\sim$1M tokens  & hosted & 241.00 s & \$0.3449 \\
    qwen3-8b                & 8B               & $\sim$32K tokens & local  & 131.24 s & n/a \\
    qwen3-32b               & 32B              & $\sim$28K tokens & local  & 400.61 s & n/a \\
    \bottomrule
    \end{tabular}
\end{table}

\subsection{Sampling Settings \& Failure Rates}
For each document and model configuration, the pipeline is executed at five decoding temperatures
$T \in \{0, 0.5, 1.0, 1.5, 2.0\}$.
Each model--temperature configuration is repeated ten times to estimate run-to-run variability under stochastic inference.
Failure rates are outlined in Table~\ref{tab:consistency-failure-rate}.

\newpage
\subsection{Workflow Graph Reconstruction Analysis}

\paragraph{Measured Quantities.}
Structural consistency is quantified using:
\begin{itemize}
    \item variability of total node counts $V$,
    \item variability of stage-specific node counts
    $(V_{\text{sources}}, V_{\text{methods}}, V_{\text{experiments}}, V_{\text{sinks}})$,
    \item variability of edge counts $E$,
    \item normalized graph edit distance between reconstructed workflow graphs.
\end{itemize}

\paragraph{Graph-Edit-Distance (GED).}
For each paper--model--temperature configuration, normalized graph-edit-distance is computed as the mean pairwise labeled graph edit distance across repeated runs divided by the mean graph size of the configuration.
Graph size is defined as the total number of nodes and edges.
Edit distance counts node and edge insertions or deletions with unit cost.

\paragraph{Aggregation Procedure.}
For each paper--model--temperature configuration, variability statistics are first computed across repeated runs.
These quantities are then averaged across paper--temperature conditions for each model.
This aggregation ensures that configurations with fewer successful executions do not receive disproportionate weight.

Reported table entries therefore represent mean variability across configurations, with standard deviations shown in parentheses.
The column \texttt{n\_runs} denotes the number of successful workflow profiles included in the aggregation.

\paragraph{Results Across Different Temperatures.}

In addition to Table~\ref{tab:workflow-consistency-graph}, which was sampled for temperature $T=0$, here (Table~\ref{tab:workflow-consistency-graph-avg-temperature}) we further provide the same table as an average across different temperatures to highlight the robustness of the approach.

\subsection{Reproducibility Assessment Score Analysis}

In addition to Table~\ref{tab:repro-score-consistency} that was sampled for temperature $T=0$, here (Table~\ref{tab:repro-score-consistency-avg-temperature}) we further provide the same table as an average across different temperatures to highlight the robustness of the approach.

\paragraph{Evaluation-subset and context-window caveat.}

A practical limitation of the analyst pipeline is the large input context it requires: a full paper, including appendices, supplementary material, and long related-work sections, routinely exceeds the effective context window of smaller open-weight models. In our sweep this constraint forced the Qwen models to be evaluated on a 7-paper subset of the corpus rather than the full 10-paper set used for the Gemini and GPT models; this is why the per-model run counts in Table~\ref{tab:consistency-failure-rate} differ (the 5.7\% failure rate observed for Qwen reflects 66 successful runs over the reduced subset, not over the  corpus). 

\clearpage

\begin{table}[!h]
\centering
\caption{\textbf{Consistency Failure Rate Across Models and Sampling Temperatures.}}
\label{tab:consistency-failure-rate}
\begin{tabular}{lccccccc}
\toprule
 & \multicolumn{7}{c}{\textbf{Temperature}} \\
\cmidrule(lr){2-8}
\textbf{Model} & \textbf{0} & \textbf{0.25} & \textbf{0.5} & \textbf{0.75} & \textbf{1} & \textbf{1.5} & \textbf{2} \\
\midrule
gemini-2.5-flash        & 23.0\% & -- & 19.0\% & -- & 19.0\% & 41.0\% & 32.0\% \\
gemini-2.5-pro          & 9.0\%  & -- & 5.0\%  & -- & 6.0\%  & 11.0\% & 5.0\% \\
gemini-3-flash-prev.    & 6.0\%  & -- & 1.0\%  & -- & 2.0\%  & 5.0\%  & 3.0\% \\
gemini-3.1-pro-prev.    & 4.0\%  & -- & 1.0\%  & -- & 0.0\%  & 0.0\%  & 4.0\% \\
gpt-4.1                 & 0.0\%  & -- & 0.0\%  & -- & 0.0\%  & 0.0\%  & 0.0\% \\
qwen3-32b               & 40.0\% & 40.0\% & 40.0\% & 40.0\% & 40.0\% & -- & -- \\
qwen3-8b                & 5.7\%  & 15.7\% & 11.7\% & 1.7\%  & 3.3\%  & -- & -- \\
\bottomrule
\end{tabular}
\end{table}

\begin{table}[!ht]
\centering
\caption{\textbf{Workflow Graph Consistency For Different Models (Across Temperatures)}}
\label{tab:workflow-consistency-graph-avg-temperature}
\begin{tabular}{lccccccc}
\toprule
 & & \multicolumn{6}{c}{\textbf{Node and Edge Count Variability}} \\
\cmidrule(lr){3-8}
\textbf{LLM (n runs)} & \textbf{GED} & {$E$} & {$V$} & {$V_{\text{sources}}$} & {$V_{\text{methods}}$} & {$V_{\text{experiments}}$} & {$V_{\text{sinks}}$} \\
\midrule
gemini-2.5-flash (366)       & 1.04 & 15.25 & 6.47 & 1.02 & 3.52 & 3.14 & 1.46 \\
gemini-2.5-pro (464)         & 0.83 & 3.39  & 2.22 & 0.65 & 1.35 & 1.19 & 0.75 \\
gemini-3-flash-preview (483) & 0.59 & 2.29  & 1.10 & 0.35 & 0.67 & 0.69 & 0.46 \\
gemini-3.1-pro-preview (491) & 0.59 & 2.99  & 1.53 & 0.14 & 0.80 & 0.88 & 0.64 \\
gpt-4.1 (500)                & 0.61 & 3.15  & 1.74 & 0.48 & 1.02 & 0.82 & 0.52 \\
qwen3-32b (300)              & 0.76 & 5.76  & 4.09 & 0.51 & 2.19 & 2.16 & 1.41 \\
qwen3-8b (295)               & 0.51 & 3.58  & 2.47 & 0.71 & 0.86 & 1.01 & 1.43 \\
\bottomrule
\end{tabular}
\end{table}

\begin{table}[!ht]
\centering
\caption{\textbf{Reproducibility Score Consistency For Different Models (Across Temperatures)}}
\label{tab:repro-score-consistency-avg-temperature}
\begin{tabular}{lcccccccc}
\toprule
& \multicolumn{8}{c}{\textbf{Reproducibility Assessment Score Variability}} \\
\cmidrule(lr){2-9}
& \multicolumn{5}{c}{\textbf{Micro-Level Assessment}} 
& \multicolumn{3}{c}{\textbf{Composite}} \\
\cmidrule(lr){2-6} \cmidrule(lr){7-9}
\textbf{LLM (n runs)} 
& \textbf{All} 
& {$V_{\text{sources}}$} 
& {$V_{\text{methods}}$} 
& {$V_{\text{experiments}}$} 
& {$V_{\text{sinks}}$} 
& {$R_c$} 
& {$R_s$} 
& {$R$} \\
\midrule
gemini-2.5-flash (366)       & 0.08 & 0.12 & 0.09 & 0.11 & 0.07 & 0.06 & 0.14 & 0.08 \\
gemini-2.5-pro (464)         & 0.08 & 0.05 & 0.14 & 0.12 & 0.10 & 0.06 & 0.13 & 0.07 \\
gemini-3-flash-preview (483) & 0.05 & 0.10 & 0.06 & 0.05 & 0.04 & 0.05 & 0.10 & 0.06 \\
gemini-3.1-pro-preview (491) & 0.05 & 0.03 & 0.09 & 0.07 & 0.04 & 0.04 & 0.07 & 0.04 \\
gpt-4.1 (500)                & 0.07 & 0.05 & 0.11 & 0.13 & 0.06 & 0.01 & 0.11 & 0.08 \\
qwen3-32b (300)              & 0.10 & 0.23 & 0.14 & 0.14 & 0.08 & 0.09 & 0.17 & 0.12 \\
qwen3-8b (295)               & 0.10 & 0.12 & 0.20 & 0.20 & 0.08 & 0.09 & 0.13 & 0.09 \\
\bottomrule
\end{tabular}
\end{table}

\clearpage
\newpage
\section*{NeurIPS Paper Checklist}

\begin{enumerate}

\item {\bf Claims}
    \item[] Question: Do the main claims made in the abstract and introduction accurately reflect the paper's contributions and scope?
    \item[] Answer: \answerYes{} 
    \item[] Justification: The abstract and introduction discuss the claims proved in the paper.

\item {\bf Limitations}
    \item[] Question: Does the paper discuss the limitations of the work performed by the authors?
    \item[] Answer: \answerYes{} 
    \item[] Justification: The paper discusses settings where the results do not apply, did not perform best, critically assesses limitations (e.g. context window size, runtime, costs, failure rate for different LLM models), and also presents negative results.

\item {\bf Theory assumptions and proofs}
    \item[] Question: For each theoretical result, does the paper provide the full set of assumptions and a complete (and correct) proof?
    \item[] Answer: \answerNA{} 
    \item[] Justification: We do not provide theoretical proofs in this work (not relevant). We provide a primarily empirical work in this submission.

    \item {\bf Experimental result reproducibility}
    \item[] Question: Does the paper fully disclose all the information needed to reproduce the main experimental results of the paper to the extent that it affects the main claims and/or conclusions of the paper (regardless of whether the code and data are provided or not)?
    \item[] Answer: \answerYes{} 
    \item[] Justification: Yes, the paper includes all relevant information needed to reproduce the main experimental results of the paper. Besides details in the main text and an extensive appendix with details, we also provide a fully open-source implementation (anonymized GitHub repository) with technical documentation submitted alongside the manuscript on OpenReview.

\item {\bf Open access to data and code}
    \item[] Question: Does the paper provide open access to the data and code, with sufficient instructions to faithfully reproduce the main experimental results, as described in supplemental material?
    \item[] Answer: \answerYes{} 
    \item[] Justification: The work leverages two publicly available datasets, and further also provides a new dataset derived from the\textit{Rescience C} journal, which is available online (open access). Meta-data of the selected 213 papers (DOIs) can be found in the extensive code zip file submitted alongside the manuscript. 

\item {\bf Experimental setting/details}
    \item[] Question: Does the paper specify all the training and test details (e.g., data splits, hyperparameter, how they were chosen, type of optimizer) necessary to understand the results?
    \item[] Answer: \answerYes{} 
    \item[] Justification: In our computational experiments, we do not execute training of models, rather we execute pipelines and use publicly available models (via API, namely Gemini and GPT) for inference. Still, we provide detailed instructions, and also the full code for transparency in the extensive code zip file submitted alongside the manuscript. 

\item {\bf Experiment statistical significance}
    \item[] Question: Does the paper report error bars suitably and correctly defined or other appropriate information about the statistical significance of the experiments?
    \item[] Answer: \answerYes{} 
    \item[] Justification: We run our pipelines with multiple runs (ten) to assess mean and standard deviation. These are reported in the consistency checks in Tables 1-4. Furthermore, Figure 2 shows statistical, distributive properties of various scores examined. Furthermore, all scripts to generate tables and figures, and all data (json files from the pipeline runs) are available in the extensive code zip file submitted alongside the manuscript. 

\item {\bf Experiments compute resources}
    \item[] Question: For each experiment, does the paper provide sufficient information on the computer resources (type of compute workers, memory, time of execution) needed to reproduce the experiments?
    \item[] Answer: \answerYes{} 
    \item[] Justification: The implementation details in the paper, and the code’s documentation (README.md) state the computing resources used. Furthermore, as we access online resources (LLMs from Gemini and GPT4) we also provide details on model architecture, runtime, and run-costs.
    
\item {\bf Code of ethics}
    \item[] Question: Does the research conducted in the paper conform, in every respect, with the NeurIPS Code of Ethics \url{https://neurips.cc/public/EthicsGuidelines}?
    \item[] Answer: \answerYes{} 
    \item[] Justification: We have reviewed and followed the NeurIPS Code of Ethics.

\item {\bf Broader impacts}
    \item[] Question: Does the paper discuss both potential positive societal impacts and negative societal impacts of the work performed?
    \item[] Answer: \answerYes{} 
    \item[] Justification: We discuss the potential for automated reproducibility assessment as complement to increasingly overwhelmed human peer review processes. At the same time, we critically discuss (based on our literature review) implications on the social contract of peer-review, and the increasing usage of LLMs by human reviewers.
    
\item {\bf Safeguards}
    \item[] Question: Does the paper describe safeguards that have been put in place for responsible release of data or models that have a high risk for misuse (e.g., pre-trained language models, image generators, or scraped datasets)?
    \item[] Answer: \answerNA{} 
    \item[] Justification: Our method and released dataset do not have a risk of misuse.

\item {\bf Licenses for existing assets}
    \item[] Question: Are the creators or original owners of assets (e.g., code, data, models), used in the paper, properly credited and are the license and terms of use explicitly mentioned and properly respected?
    \item[] Answer: \answerYes{} 
    \item[] Justification: The licsenses and terms of use for all the pre-existing relevant data, code and models used in the benchmark (Reproscreener, Repro-Bench) were respected.

\item {\bf New assets}
    \item[] Question: Are new assets introduced in the paper well documented and is the documentation provided alongside the assets?
    \item[] Answer: \answerYes{} 
    \item[] Justification: We have properly documented our repository via ReadMe.md files; including implementation details, and setup and execution instructions. Furthermore, there is an extensive appendix to the main paper.

\item {\bf Crowdsourcing and research with human subjects}
    \item[] Question: For crowdsourcing experiments and research with human subjects, does the paper include the full text of instructions given to participants and screenshots, if applicable, as well as details about compensation (if any)? 
    \item[] Answer: \answerNA{} 
    \item[] Justification: The contributions of the paper do not relate to human subjects.

\item {\bf Institutional review board (IRB) approvals or equivalent for research with human subjects}
    \item[] Question: Does the paper describe potential risks incurred by study participants, whether such risks were disclosed to the subjects, and whether Institutional Review Board (IRB) approvals (or an equivalent approval/review based on the requirements of your country or institution) were obtained?
    \item[] Answer: \answerNA{} 
    \item[] Justification: This does not apply to our research (not necessary, applicable).

\item {\bf Declaration of LLM usage}
    \item[] Question: Does the paper describe the usage of LLMs if it is an important, original, or non-standard component of the core methods in this research? Note that if the LLM is used only for writing, editing, or formatting purposes and does \emph{not} impact the core methodology, scientific rigor, or originality of the research, declaration is not required.
    \item[] Answer: \answerYes{} 
    \item[] Justification: LLMs are substantial components of our proposed agentic pipeline. We clearly document (for scientific rigorousness) model architectures, prompts and run executions used, which can be found in the manuscript, its appendix, and the extensive code appendix which is submitted alongside the manuscript on OpenReview. To not only rely on commercial models (Gemini, GPT) we also run experiments on open source models on our own GPU facilities. Prompts were cautiously designed over many iterations, and extensive, systematic consistency assessments show that our pipeline finds meaningful patterns rather than to hallucinate.

\end{enumerate}

\end{document}